\documentclass[journal,10pt]{IEEEtran}

\usepackage{graphicx}
\usepackage{float}
\usepackage{bm}

\usepackage{cite}

\usepackage{xcolor}      
\usepackage[
  colorlinks   = true,     
  linkcolor    = blue,       
  citecolor    = blue,        
  urlcolor     = blue,        
  filecolor    = blue        
]{hyperref}

\let\oldcite\cite
\renewcommand{\cite}[1]{\textcolor{blue}{\oldcite{#1}}}

\usepackage{subfigure}
\usepackage{amsmath,amsfonts}
\usepackage{mathtools}
\usepackage{algorithm}
\usepackage{algpseudocode}
\usepackage[normalem]{ulem}
\usepackage[tableposition=top]{caption}
\usepackage{url}
\usepackage{booktabs}

\usepackage{array}
\usepackage{color}
\usepackage{cleveref}
\usepackage{geometry}
\usepackage{amssymb}
\usepackage{cuted}
\usepackage{lipsum} 
\usepackage{multicol}

\renewcommand{\eqref}[1]{\textcolor{blue}{(}\ref{#1}\textcolor{blue}{)}}

\usepackage{geometry}
\begin{document}

\title{Trellis Waveform Shaping for Sidelobe Reduction in Integrated Sensing and Communications: \\A Duality with PAPR Mitigation}

\author{
Henglin Pu,
Husheng Li, \IEEEmembership{Senior Member, IEEE},
Zhu Han, \IEEEmembership{Fellow, IEEE},
H. Vincent Poor, \IEEEmembership{Life Fellow, IEEE}


\thanks{H. Pu is with the Elmore Family School of Electrical and Computer Engineering, Purdue University, West Lafayette, Indiana, USA-47907 (email:
 pu36@purdue.edu).}%
\and
\thanks{Z. Han is with the Department of Electrical Engineering, University of Houston (email: zhan2@uh.edu).} \and\thanks{H. V. Poor is with the Department of Electrical and
Computer Engineering, Princeton University (email: poor@princeton.edu).}\and
\thanks{H. Li is with the School of Aeronautics and Astronautics, and the Elmore Family School of Electrical and Computer Engineering, Purdue University, USA (email: husheng@purdue.edu). (email: husheng@purdue.edu).}\and\thanks{This work was supported by the National Science Foundation under grants 2348826, 2343465, 2343469 and 2418106. }
}

\maketitle

\begin{abstract}
A key challenge in integrated sensing and communications (ISAC) is the synthesis of waveforms that can modulate communication messages and achieve good sensing performance simultaneously. In ISAC systems, standard communication waveforms can be adapted for sensing, as the sensing receiver (co-located with the transmitter) has knowledge of the communication message and consequently the waveform. However, the randomness of communications may result in waveforms that have high sidelobes masking weak targets. Thus, it is desirable to refine communication waveforms to improve the sensing performance by reducing the integrated sidelobe levels (ISL). This is similar to the peak-to-average power ratio (PAPR) mitigation in orthogonal frequency division multiplexing (OFDM), in which the OFDM-modulated waveform needs to be refined to reduce the PAPR. In this paper, inspired by PAPR reduction algorithms in OFDM, we employ trellis shaping in OFDM-based ISAC systems to refine waveforms for specific sensing metrics using convolutional codes and Viterbi decoding. In such a scheme, the communication data is encoded and then mapped to the signaling constellation in different subcarriers, such that the time-domain sidelobes are reduced. An interesting observation is that sidelobe reduction in OFDM-based ISAC is dual to PAPR reduction in OFDM, thereby sharing a similar signaling structure. Numerical simulations and hardware software defined radio USRP experiments are carried out to demonstrate the effectiveness of the proposed trellis shaping approach. 
\end{abstract}


\section{Introduction}
Integrated sensing and communications (ISAC) integrates both functions in the same waveform, thus substantially improving the spectral and power efficiencies. It is expected to be a distinguishing feature of 6G wireless communication networks. In ISAC, the forward propagation of electromagnetic (EM) wave “pushes" communication messages to the destination, thereby accomplishing the task of communications, and “pulls" the environmental information in the backward propagation upon reflections and scatterering, thus achieving the function of sensing. 

The major challenge of ISAC is the synthesis of waveforms satisfying both communications and sensing performance goals. Three possible design criteria can be employed: 
\begin{enumerate}
    \item Communication-centric design, where the waveform synthesis stems from an existing communication waveform (such as the orthogonal frequency division multiplexing (OFDM) and sensing is accomplished by considering the communication waveform as a pseudorandom sequence. The advantage of such schemes is that the communication performance can be well assured, while the disadvantage is that the random communication messages cause uncertainties in the sensing performance.
    \item Sensing-centric design, where communication messages are embedded in the traditional radar sensing waveforms (e.g. FMCW), such as modulating the radar parameters (e.g., the chirp rate)\cite{HLi_Globecom_2019_radar} or using spread spectrum (either direct sequence\cite{LiJCS2023} or frequency hopping patterns\cite{AboutaniosRadarCon2023}). Such schemes are convenient but difficult to achieve high communication data rates due to the lack of an explicit communication signaling structure.
    \item Dedicated ISAC waveforms, which are designed from first principles and optimized for achieving a good performance trade-off between communications and sensing, at the cost of greater design and/or computational complexity.
\end{enumerate}


\begin{figure}[!t]
\centering
\includegraphics[width=0.4\textwidth]{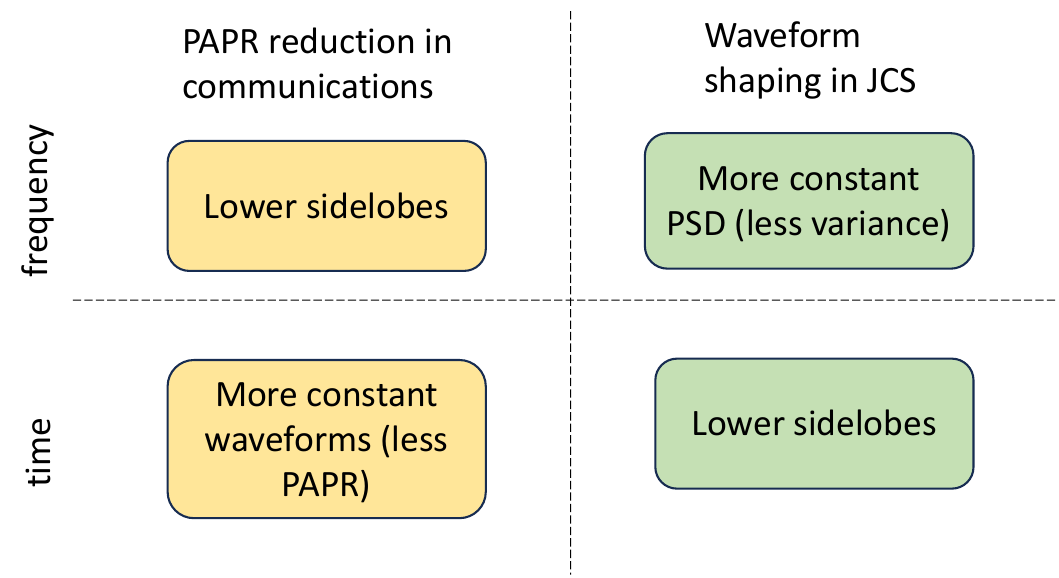}
\caption{Conceptual duality between OFDM PAPR mitigation (left) and ISAC waveform shaping (right). Flattening the time‑domain envelope of an OFDM symbol lowers PAPR, while flattening the frequency‑domain PSD of the same symbol lowers the ISL}
\label{fig:duality}
\end{figure}

To meet those design criteria in this paper, we adopt the communication-centric design philosophy and assume the OFDM signaling. On the one hand, the popular communication signal structure of OFDM can guarantee the communication performance and reuse existing communication system hardware and protocols; on the other hand, existing communication waveforms, without any modifications, can indeed be used for sensing (e.g., the extensive work on the Wi-Fi sensing\cite{wifi_sensing,wifi_sensing2}). However, as indicated above, pure communication waveforms may bring significant performance degradations to sensing. In particular, the random OFDM waveform may result in significant sidelobes that mask weak radar targets. (See\cite{ureten2009autocorrelation} for an example). 


To address this challenge in communication-centric ISAC, we leverage the lessons learned from mitigating the peak-to-average power ratio (PAPR) in OFDM systems, which has received extensive study over the past two decades and has been employed in 4G and 5G systems. It is well known that one of the major disadvantages of multi-carrier communications is that the peak transmit power can be substantially higher than the average power, thus bringing potential distortions to the power amplifier of the radio frequency (RF) circuits. There are many algorithms for PAPR reduction such as the peak clipping\cite{Neil1994}, selective mapping\cite{Eetvelt1996}, trellis shaping\cite{Ochiai2004}, and so on. A survey can be found in\cite{Litsyn2007}. Here, a high PAPR can be considered as a byproduct of communication modulation. Similarly, the sensing performance degradation (such as the greater sidelobe level) of ISAC, compared with traditional radar sensing waveforms, can also be considered as a byproduct of communication modulation. Therefore, it is possible that approaches for alleviating the PAPR in OFDM waveforms can be leveraged (not necessarily in all details) for improving the sensing performance in communication-centric ISAC. 

In this paper, we propose to adopt the approach of trellis shaping\cite{Ochiai2004} that has been employed for PAPR mitigation, as well as power reduction. The idea is similar to coset coding\cite{Forney1988}, in which the communication message is modulated in the index of a coset (i.e., a subset of codewords). Within the coset, the codewords having the optimal performance (e.g., the minimal PAPR or the least sidelobes) will be selected for transmissions. The larger the coset is, the better sensing performance can be achieved since there are more options, while a reduction in the data rate is certainly incurred. In the context of ISAC, we will focus on minimizing the sidelobes in the waveform autocorrelation, in order to avoid confusion between sidelobes and weak targets. The goal of sidelobe reduction in ISAC is dual to the PAPR mitigation in OFDM, since in ISAC we expect to reduce the sidelobes in the time domain while for PAPR we desire to mitigate the sidelobes in the frequency domain (thus a flatter power profile in the time domain). As illustrated in Fig.~\ref{fig:duality}, flattening the OFDM envelope for PAPR mitigation in conventional communications and flattening the power‑spectral density for sidelobe suppression in ISAC are two sides of the same Fourier‑domain coin. The challenge to our waveform shaping in ISAC is that the manipulation of the signal is in the frequency domain due to the OFDM structure while the goal is in the time domain, which makes it more challenging than trellis-shaping-based PAPR reduction. Note that the possible solutions to waveform design in ISAC are not limited to trellis shaping. The more important lesson is that the experience learned in PAPR mitigation that has been intensively studied for OFDM communication systems in the past decades can be leveraged in the context of ISAC. Many other methodologies in PAPR mitigation\cite{Neil1994,Eetvelt1996,Ochiai2004} may also be exploited for this purpose. Simply put, our main contributions can be summarized as
follows:

\begin{itemize}
\item We re‑cast sidelobe suppression in OFDM‑ISAC as the time‑domain dual of PAPR mitigation and proposes a trellis‑shaping encoder to refine subcarrier powers accordingly.

\item We introduce a two‑regime cost-PSD variance for long symbols, explicit ISL for short ones—and embed it in a per‑subcarrier Viterbi recursion, enabling sign‑bit or multidimensional shaping with tunable redundancy.

\item We formulate a normalised weighted cost that stabilises decoding and quantifies the inherent ISL–PAPR trade‑off, yielding a single knob to navigate sensing vs. amplifier efficiency.

\item Extensive simulations (16‑/256‑QAM, 32–1024 subcarriers) show that the proposed shaping achieves $10$–$50\%$ ISL reduction and comparable PAPR relief, with controllable redundancy (e.g.\ $3/4$ symbol rate for sign‑bit shaping) and acceptable BER penalties.

\item An mmWave SDR testbed confirms that shaped waveforms maintain lower sidelobes and reduced PAPR after over‑the‑air transmission, and improve monostatic radar ranging—demonstrating real‑world feasibility of trellis‑based ISAC waveform synthesis.
\end{itemize}

 The remainder of this paper is organized as follows. Related studies are briefly introduced in Section \ref{sec:related}. Then, the system model and preliminary of trellis shaping are introduced in Section \ref{sec:model} and Section \ref{sec:prelim} separately. The main part of this paper, namely the algorithm for waveform shaping and the trade-off between communications and sensing, is contained in Sections \ref{sec:ISL_shaping} and \ref{sec:joint}. Then, numerical and experimental results are provided in Section \ref{sec:numerical}, and final conclusions are drawn in Section \ref{sec:conclusions}. 

\section{Related Works}\label{sec:related}

There has been considerable interest in ISAC in recent years. Surveys on ISAC can be found in\cite{Zheng2019,FanLiu2020,Ma2020}. The spectrum of ISAC waveform designs ranges from using traditional radar sensing waveforms (e.g. FMCW\cite{HLi_Globecom_2019_radar}) to traditional communication waveforms (e.g., Wi-Fi waveforms\cite{wifi_sensing,wifi_sensing2}). There are many studies of using pure communication signals for radar sensing\cite{wifi_sensing,4977002,9248648}, while there are fewer studies on refining communication signals in order to improve the sensing performance.

As highlighted in the introduction, the challenge of reducing sidelobes in ISAC waveforms parallels the PAPR mitigation problem in OFDM systems. In OFDM, high PAPR compromises the efficiency of high-power amplifiers (HPAs)\cite{wunder2013papr}, spurring research into techniques including clipping and filtering techniques\cite{wang2010optimized,armstrong2002peak}, active constellation extension (ACE)\cite{krongold2003reduction,sohn2014low}, and tone reservation (TR) methods\cite{li2011improved,wang2008analysis}. While effective for communication, these methods have seen limited adaptation for sensing, where waveform design must also minimize autocorrelation sidelobes for enhanced target detection\cite{stoica2009new,he2009designing,stoica2009designing}. This dual focus on low autocorrelation sidelobes (e.g., integrated/peak sidelobe levels, ISL/PSL) and low PAPR has driven advancements in ISAC systems, particularly those employing constant modulus waveforms\cite{shi2019spectrally,song2015optimization,jing2018designing}. However, the constant modulus constraint is often overly stringent and lacks flexibility, which can be detrimental to system performance. 

The simultaneous minimization of the PAPR and autocorrelation sidelobes in OFDM signals has emerged as a critical research frontier, particularly for applications requiring dual radar-communication functionality. Recent research addressing the joint challenge of PAPR and time-domain sidelobe suppression in OFDM radar leverages both classical signal-processing solutions and more advanced, often data-driven, frameworks. Early work focused on shaping the transmit waveform through windowing, filtering, or subcarrier weighting to curtail autocorrelation sidelobes and limit large amplitude peaks simultaneously\cite{zhou2013active}. More recent studies formulate multi-objective or constrained optimization frameworks, employing either convex programming or iterative approaches to enforce radar sidelobe constraints alongside stringent PAPR limits\cite{raghavendra2020joint,feng2024sidelobe}. In parallel, deep learning and autoencoder-based methods have emerged to automatically discover low-PAPR waveforms with improved correlation properties by jointly training with respect to radar performance indicators\cite{liu2023radar,xia2024novel}.

However, existing ISAC waveform research either (i) imposes a constant‑envelope/PAPR cap and then suppresses autocorrelation sidelobes through windowing, convex optimisation, or deep‑learning–based auto‑encoders, or (ii) tackles PAPR and ISL separately, rarely sharing design machinery across the two objectives. In contrast, our work is the first to transplant trellis shaping, a method originally devised for PAPR control in OFDM, into the ISAC context. We further unify ISL and PAPR in a single normalized Viterbi metric by exploiting the duality between PAPR mitigation and sidelobe reduction, yielding linear‑complexity dynamic programming instead of iterative solvers. A preliminary conference version of this study was presented in\cite{conf_ver}. That paper introduced the baseline trellis-shaping concept for lowering the ISL in OFDM-based ISAC waveforms and validated it through numerical simulations.

\section{OFDM System and Metrics}\label{sec:model}
In this study, we consider an OFDM symbol $\left\{ X_n \right\}_{n=0}^{N_s-1}$ comprising $N_s$ subcarriers, each transmitting an average of $N_b$ bits. The time-domain signal samples are denoted by $\{ x_n \}_{n=0, \ldots, N_s-1}$, with a total transmit power of $P_t$, i.e., 
\begin{equation}\label{eq:OFDM_sig}
x_n = \frac{1}{\sqrt{N_s}} \sum_{k=0}^{N_s-1} X_{k} e^{j 2\pi kn / N_s}, \quad n = 0, \ldots, N_s - 1.
\end{equation}

The PAPR is a key metric that describes the dynamic range of the OFDM signals in the time domain. The PARR can be used to estimate how much back-off is needed for the transmit power in order to avoid nonlinear amplification or clipping at the transmitter. Considering the PAPR as a random variable, the conventional definition of the PAPR for the OFDM symbol is given by
\begin{equation}
\text{PAPR} = \frac{\max_n |x_n|^2}{E[\,|x_n|^2\,]}\,.
\end{equation}

For the purpose of sensing, we define the aperiodic time-domain autocorrelation function as
\begin{eqnarray}
r(l)=\sum_{n=0}^{N_s-l}x_n^*x_{n+l},\qquad l=0,...,N_s-1,
\end{eqnarray}
which quantifies the signal's self-similarity at different lags. Sensing performance is assessed via the normalized ISL, which is defined as
\begin{eqnarray}\label{eq:ISL}
\text{ISL}=\frac{\sum_{k=1}^{N_s-1}|r(l)|^2}{|r(0)|^2},
\end{eqnarray}
where $|r(0)|^2$ is the mainlobe power, and the goal is to minimize sidelobe amplitudes $|r(l)|$ for $l > 0$ to enhance the sensing accuracy by reducing interference with weak targets.

For infinite-length signals, the autocorrelation function and power spectral density (PSD) are Fourier transform pairs. A flat PSD, with power evenly distributed across frequencies, yields an autocorrelation function approximating a Dirac delta function with minimal sidelobes. In our finite-length OFDM system with $N_s$ subcarriers, the spectrum is discrete, and the PSD is approximated by $|X[k]|^2$, the squared magnitude of the discrete Fourier transform (DFT) of $x[k]$.

To justify flattening the PSD in the discrete frequency spectrum to reduce sidelobes, we analyze the circular autocorrelation of the finite sequence
\begin{equation}
   r_c(l) = \sum_{n=0}^{N_s-1} x_n^* x_{(n + l)\bmod N_s}, \quad l = 0, \ldots, N_s - 1. 
\end{equation}

According to the Wiener-Khinchin theorem\cite{wiener1930generalized}, the DFT of $r_c(l)$ yields
\begin{equation}
   \text{DFT}\{ r_c(l) \} = |X_k|^2, 
\end{equation}
showing that a uniform $|X[k]|^2$ makes $r_c(l)$ approximate a Kronecker delta, namely $r_c(l) \approx c \cdot \delta[l]$, thus minimizing sidelobes. In sensing, however, the aperiodic $r(l)$ is relevant, corresponding to matched filter outputs.

\begin{figure}[!t]
\centering
\includegraphics[width=0.38\textwidth]{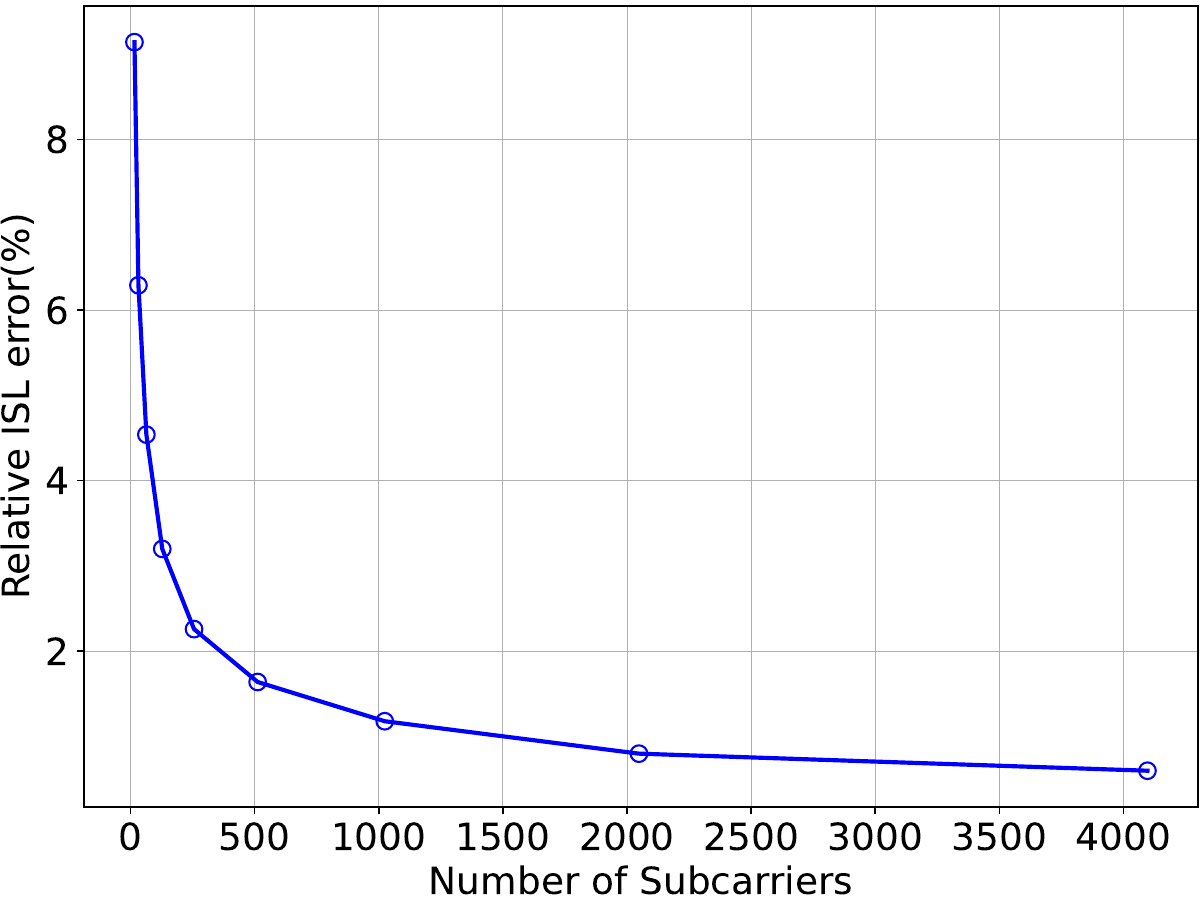}
\caption{ISL error between aperiodic correlation and periodic correlation versus different numbers of subcarriers. Above $N_s\approx 256$ the error falls below $2\%$, justifying the PSD‑variance surrogate used for large‑bandwidth systems.}
\label{fig:ISL_difference}
\end{figure}

For small lags $l \ll N_s$, especially with large $N_s$, $r(l) \approx r_c(l)$, as the periodic wrapping in $r_c(l)$ has a minimal effect. The difference is given by
\begin{equation}
  r_c(l) - r(l) = \sum_{k=N_s - l + 1}^{N_s} x_n^* x_{n + l - N_s},  
\end{equation}
representing wrapped terms. The approximation $r(l) \approx r_c(l)$ introduces a bias, $r_c(l) - r(l)$, exact for infinite signals but imperfect for finite $N_s$. The bias magnitude is given by
\begin{equation}
   |r_c(l) - r(l)| = \left| \sum_{n=N_s - l + 1}^{N_s} x_n^* x_{n + l - N_s} \right|. 
\end{equation}

Using the Cauchy-Schwarz inequality, we have
\begin{equation}
   |r_c(l) - r(l)| \leq \sqrt{ \sum_{k=N_s - l + 1}^{N_s} |x_n|^2 \sum_{k=N_s - l + 1}^{N_s} |x_{n + l - N_s}|^2 }.
\end{equation}
Assuming a uniform power $|x_n|^2 \approx P_t / N_s$, we have
\begin{equation}
    |r_c(l) - r(l)| \leq \frac{l P_t}{N_s}.
\end{equation}

This bias is small relative to $r(0) = P_t$ for $l \ll N_s$. It grows with $l$, potentially reaching $P_t/2$ for $l \approx N_s/2$, and thus affecting the ISL if sidelobes are not minimized.

For random data (e.g., QAM symbols), the expected bias $E[r_c(l) - r(l)] = 0$, as $E[x_n^* x_{n + l - N_s}] = 0$. However, the variance
\begin{equation}
  \text{Var}[r_c(l) - r(l)] = E\left[ \left| \sum_{k=N_s - l + 1}^{N_s} x_n^* x_{n + l - N_s} \right|^2 \right],  
\end{equation}
scales with $l \cdot (P_t / N_s)^2$, impacting individual ISL realizations. For a large $N_s$, normalization by $|r(0)|^2 \approx P_t^2$ mitigates this, thus making a flat PSD a practical approach to minimize expected ISL. In Fig.~\ref{fig:ISL_difference}, we compare the calculated relative ISL error between the aperiodic correlation result (related to the PSD) and the periodic correlation result for various numbers of subcarriers. It can be observed that for a large $N_s$ (e.g., $N_s \ge 256$), the ISL error is reduced to within 2\%.

\section{Preliminary of Trellis Shaping}\label{sec:prelim}

In this section, we briefly introduce the principle of trellis shaping for signal performance improvement in terms of PAPR reduction. The approach of trellis shaping was originally proposed by Forney in order to reduce the signal transmit power\cite{Forney1992}, but it finds further applications in the PAPR reduction\cite{Ochiai2004}. We consider $N_s$ subcarriers, each carrying QAM modulation. The time-domain signal is given by \eqref{eq:OFDM_sig}.
The PAPR is determined by the power profile in the time-domain. As we have discussed, the time-domain (respectively the frequency-domain) power profile and the autocorrelation of the frequency spectrum (respectively the time domain) form a Fourier transform pair. Therefore, reducing PAPR is equivalent to reducing the sidelobes of the frequency spectrum autocorrelation defined as
\begin{eqnarray}\label{eq:autocorrelation_F}
R(l)=\sum_{k=1}^{N_s-l}X_k^*X_{k+l}.
\end{eqnarray}
Now, assume that each subcarrier has $N_b$ bits, denoted by $\{b_{k1},...,b_{k,N_b}\}$ for subcarrier $k$, to transmit. We consider $M-1$ bits out of the $N_b$ bits to be the most significant bits (MSBs) that have the greatest impact on the waveform and consider the remaining $N_b-M$ bits as the least significant bits (LSBs) that have minor impact on the waveform. MSBs will be encoded for waveform shaping while the LSBs will be transmitted as they are. We consider expanding (encoding) the $M$ MSBs into $NM$ bits, namely for each subcarrier $k$ we choose a mapping $F_k$: $\{b_{k1},...,b_{kM}\}\rightarrow \{z_{k1},...,z_{k,MN}\}$. 

Now, we assume that $N_b=M^2-1$ and consider a convolutional code with coding rate $\frac{1}{M}$, whose generating matrix of dimension $1\times M$ is $\mathbf{G}$ (in the generating polynomial form) and whose parity check matrix of dimension $(M-1)\times M$ is $\mathbf{H}$. For example, the generating matrix could be $\mathbf{G}=(1+D^2,1+D+D^2)$, where $D$ is the unit delay. As illustrated in Fig. \ref{fig:trellis}, we take $M-1$ raw bits $\mathbf{s}$ at each subcarrier are multiplied with the inverse syndrome former matrix $(\mathbf{H}^{-1})^T$, to form $M$ MSB bits: 
\begin{eqnarray}
\mathbf{z}=\mathbf{s}(\mathbf{H}^{-1})^T. 
\end{eqnarray}
Then, $\mathbf{z}$ and the $M^2-M$ LSB bits $\mathbf{b}$ (therefore $M^2$ bits in $(\mathbf{z},\mathbf{b})$) are fed into the convolutional decoder, thus generating $M$ MSB bits $\mathbf{y}$ (as a codeword). Then, $\mathbf{z}+\mathbf{y}$ and $\mathbf{b}$ are put into the constellation mapper and sent to the radio frequency (RF) circuits for transmission. At the receiver, the MSB is multiplied by $\mathbf{H}^T$ such that
\begin{eqnarray}
(\mathbf{z}+\mathbf{y})\mathbf{H}^T=\mathbf{s}(\mathbf{H}^{-1})^T\mathbf{H}^T+\mathbf{y}\mathbf{H}^T=\mathbf{s},
\end{eqnarray}
where $\mathbf{y}\mathbf{H}^T=0$ because $\mathbf{y}$ is the outcome of the convolutional decoding (thus being a codeword). Hence, $\mathbf{s}$ is recovered at the receiver.  

To reduce PAPR, we carry out the encoding in a subcarrier-by-subcarrier manner, due to the summation structure in \eqref{eq:autocorrelation_F}, which perfectly fits the approach of dynamic programming (or Viterbi decoding in the context convolutional coding). Assume that the decoding of MSBs of the first $k-1$ subcarriers has been completed. For subcarrier $k$, the $M$ bits are decoded in the manner of Viterbi decoding. Due to the summation structure in \eqref{eq:autocorrelation_F}, the decoding metric is the power sum of sidelobes. Additional details of the trellis shaping can be found in\cite{Ochiai2004}.

\section{Trellis Waveform Shaping for ISL Reduction}\label{sec:ISL_shaping}
In this section, we employ the technique of trellis shaping in the context of ISAC, in order to reduce the time-domain autocorrelation sidelobes, similarly to reducing the PAPR in OFDM.
 
\subsection{Metric for ISL}
\begin{figure}[!t]
\centering
\includegraphics[width=0.45\textwidth]{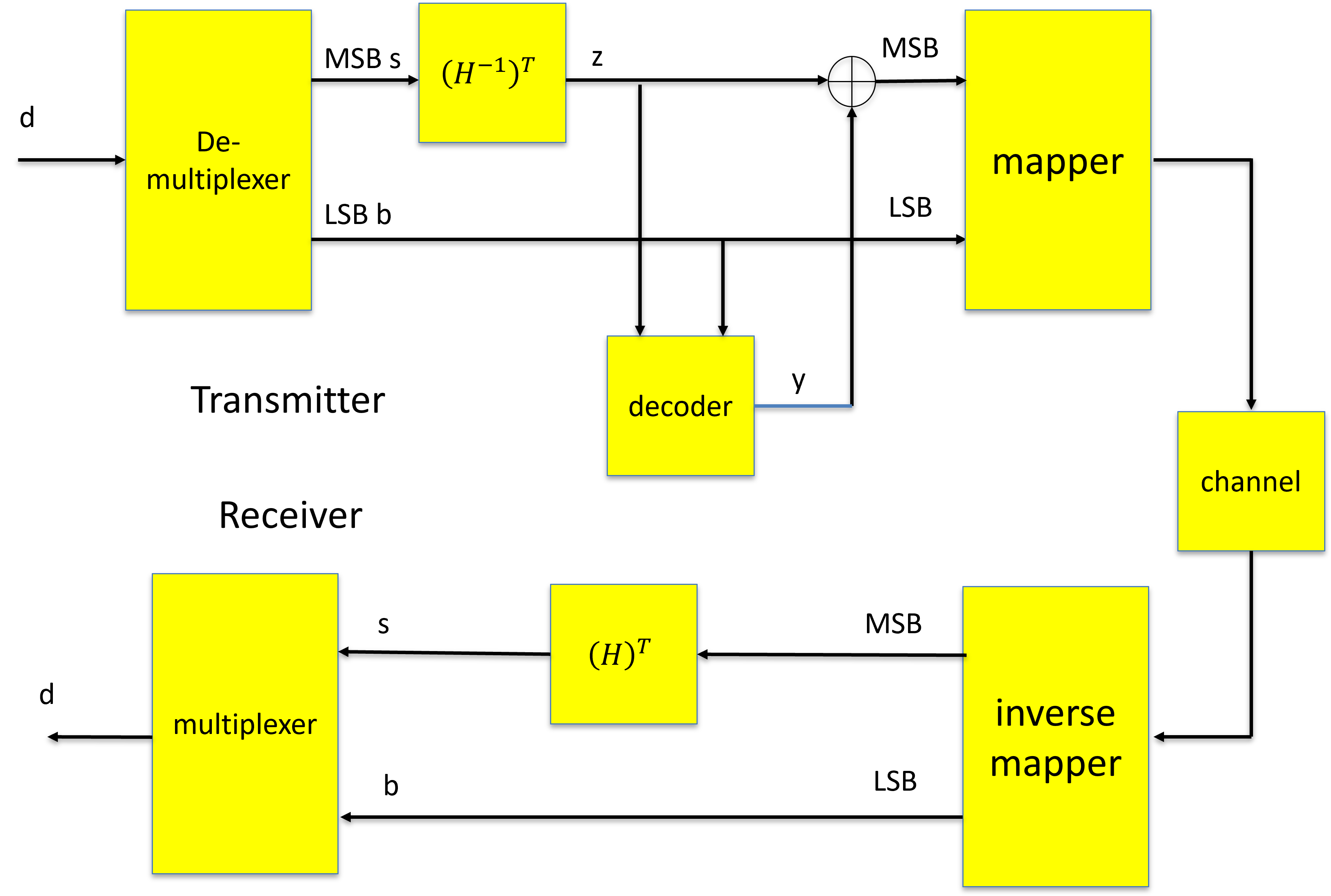}
\caption{Trellis waveform shaping}
\label{fig:trellis}
\end{figure}

We use the same structure of trellis shaping for PAPR reduction in OFDM, as shown in Fig. \ref{fig:trellis}. We fix a convolutional code with codebook $\mathcal{C}$ and data rate $\frac{1}{M}$. However, the waveform synthesis in ISAC is not completely dual to that for PAPR reduction. In ISAC, the encoding is in the frequency domain, due to the OFDM signaling structure, and the autocorrelation is in the time domain. In contrast, both encoding and autocorrelation are in the frequency domain for PAPR reduction in OFDM signaling. 

Based on the discussion in the previous section, we propose to minimize the following metric to mitigate ISL:
\begin{equation}\label{eq:ISL_metric}
    V_I(X) =
\begin{cases} 
\frac{1}{N_s} \sum\limits_{k=1}^{N_s} \left( \left| X_k \right|^2 - \frac{P_t}{N_s} \right)^2, & \text{if } N_s \geq N_0, \\[12pt]
\frac{\sum\limits_{\ell=1}^{N_s-1} \left| r(\ell) \right|^2}{\left| r(0) \right|^2}, & \text{if } N_s < N_0.
\end{cases}
\end{equation}

\noindent When $N_s$ is large ($N_s \ge N_0$, where $N_0$ is a threshold), we minimize the variance of subcarrier powers, ensuring $E[|X_k|^2] = P_t / N_s$, which promotes uniform power allocation and reduces ISL. Otherwise, the original ISL (computed using aperiodic correlation) is minimized. The advantage of this approach is that, by transitioning from aperiodic correlation to variance minimization of the PSD, the computational burden is significantly reduced. This method eliminates the need to store and compute the entire sequence length, potentially lowering the complexity of the Viterbi algorithm. The transmit power $P_t$ allows slight deviations, and trellis shaping uses a convolutional code with rate $\frac{1}{M}$, memory $\nu$, and states $\phi_1, \ldots, \phi_{2^\nu}$.

\subsection{MSB Decoding}
Given the metric in \eqref{eq:ISL_metric}, the goal of decoding, mapping from the bits $(\mathbf{z},\mathbf{b})$ to $\mathbf{y}$, is to minimize the spectrum variance, namely
\begin{eqnarray}
\mathbf{y}=\arg\min_{\mathbf{y}\in \mathcal{C}}V_I(\mathbf{y}),
\end{eqnarray}
where the approximated variance in \eqref{eq:ISL_metric} is a function of the codeword $\mathbf{y}$ since the final symbol in the constellation is determined by $\mathbf{y}+\mathbf{z}$ and $\mathbf{b}$. Then, we consider the schemes of sign-bit signaling ($M=2$) and high-dimensional signaling ($M>2$) in the sequel.

\subsubsection{Sign-bit Signaling}
For the simplest sign-bit signaling scheme, we set $M=2$. Therefore, each subcarrier has $M-1=1$ MSB bit, denoted by $s_1$, ..., $s_{N_s}$, and $M(M-1)=2$ LSB bits, denoted by $b_{11},b_{12},...,b_{N_s1},b_{N_s2}$. Note that, when the generating matrix is $\mathbf{G}=(1+D^2,1+D+D^2)$, the matrix $(\mathbf{H}^{-1})^T$ may not be unique. One possible selection is $(D,1+D)^T$. The bit vector $\mathbf{s}$ used to generate the MSB $\mathbf{z}$ is given by the following polynomial (instead of a single bit) in terms of delay $D$:
\begin{eqnarray}
\mathbf{s}(D)=\sum_{i=1}^{N_s}s_i D^{i-1}.
\end{eqnarray}
Similarly, the MSB $\mathbf{z}=\mathbf{s}\mathbf{G}$ is a vector of polynomails
consisting of $2N_s$ bits given by
\begin{eqnarray}
\mathbf{z}=\left(\underbrace{z_0,z_1}_{\text{subcarrier } 1},\underbrace{z_2,z_3}_{\text{subcarrier }2},\cdots, \underbrace{z_{2N_s-2},z_{2N_s-1}}_{\text{subcarrier }N_s}\right).
\end{eqnarray}
The corresponding polynomial form of $\mathbf{z}$ can be written as $\mathbf{z}(D)=(\mathbf{z}_1(D),\mathbf{z}_2(D))$, where $\mathbf{z}_{j}(D)=\sum_{i=1}^{N_s}z_{ji} D^{i-1}$, for $j=1,2$.

\begin{figure}[!t]
\centering
\includegraphics[width=0.35\textwidth]{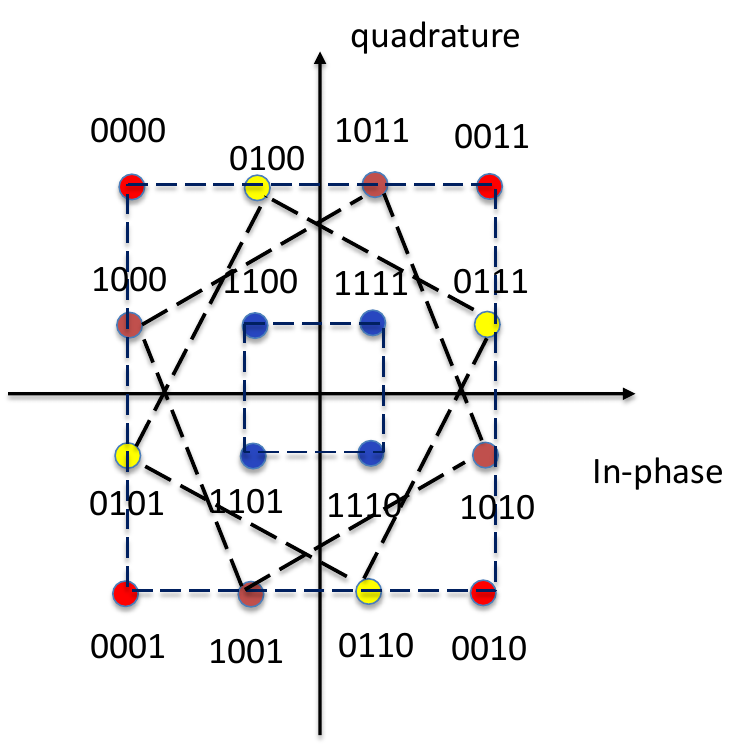}
\caption{16QAM constellation annotated for ISAC trellis shaping. The two MSB bits partition the plane into four quadrants (coloured), giving coarse control of symbol amplitude and phase; the remaining LSBs fine‑tune modulation.}
\label{fig:QAM}
\end{figure}

We can consider the MSB bits $\mathbf{s}$ as the symdrone of the bits $\mathbf{z}$. 
The bits $\mathbf{z}$ and $\mathbf{b}$, together of dimension $M^2$=4, are put into the convolutional decoder, in order to find a codeword $\mathbf{y}$ of dimension $M=2$. Moreover, the codeword $\mathbf{y}$ should make the MSB $\mathbf{z}+\mathbf{y}$ generate $N_s$ symbols (over the $N_s$ subcarriers) that yield a flat PSD (thus less sidelobes in the time-domain autocorrelation for a better sensing performance). To this end, the convolutional decoding procedure is carried out by the following Bellman equation:
\begin{equation}\label{eq:Bellman}
\begin{aligned}
        V_I(\phi_t) &=
\begin{cases} 
V_{\text{large}}(\phi_t), & \text{if } N_s \geq N_0, \\
V_{\text{small}}(\phi_t), & \text{if } N_s < N_0.
\end{cases}\\
V_{\text{large}}(\phi_t)&=\arg\min_{\phi_{t+1}\sim \phi_t} V_{\text{large}}(\phi_{t+1})+\left(|X_t|^2-\frac{P_t}{N_s}\right)^2,\\
V_{\text{small}}(\phi_{t})
&=
\arg\min_{\phi_{t+1}\,\sim\,\phi_{t}}
V_{\text{small}}(\phi_{t+1})+\Delta\bigl(\phi_{t},\,\phi_{t+1}\bigr),
\end{aligned}
\end{equation}
where $\phi_t$ is the state of the trellis for subcarrier $t$, the notion $\phi_t\sim \phi_{t+1}$ means that the two states are connected in the trellis, and $X_t$ is determined by the communication symbol determined by the transition from $\phi_t$ to $\phi_{t+1}$. Recall that in the small-$N_s$ regime, we opt to minimize the aperiodic ISL directly. Let us define
\begin{equation}
   r^{(t)}(\ell)
~=~
\sum_{n=1}^{\,t-\ell}
  x^{(t)}[n]^{*}\,x^{(t)}[n+\ell],
\quad
\ell = 1,\dots,t-1, 
\end{equation}
as the partial aperiodic correlation of the length-$t$ time-domain signal 
$x^{(t)}$. We then define
\begin{equation}
   \mu(t)
~=~
\sum_{\ell=1}^{t-1}
  \bigl\lvert\,r^{(t)}(\ell)\bigr\rvert^{2} 
\end{equation}
to be the accumulated sidelobe energy after making decisions over subcarriers $1,\dots,t$.
When adding subcarrier $t+1$, the incremental cost is given by
\begin{equation}
   \Delta\bigl(\phi_{t},\phi_{t+1}\bigr)
~=~
\mu(t+1)
~-~
\mu(t), 
\end{equation}
which measures how that new subcarrier (and thus new time-domain samples) 
contribute to the partial ISL. Hence, the Bellman recursion for small-$N_{s}$ 
is given by
\begin{equation}
V_{\text{small}}(\phi_{t})
~=~
\min_{\phi_{t+1}\,\sim\,\phi_{t}}
  V_{\text{small}}(\phi_{t+1})
  ~+~
  \Delta\bigl(\phi_{t},\,\phi_{t+1}\bigr),
\label{eq:BellmanSmall}
\end{equation}
with boundary condition $V_{\text{small}}(\phi_{N_{s}+1}) = 0$. Consequently, 
for each subcarrier $t$, the state $\phi_{t}$ evolves in the trellis by 
selecting one of the possible coded symbols, incurring an incremental 
aperiodic-sidelobe penalty. The boundary condition for the Bellman's equation is $V(s_{N_s+1})=0$. Then, the Viterbi decoding algorithm (or equivalently dynamic programming) is applied recursively to solve the equation in \eqref{eq:Bellman}. 

The convolutional decoding output is given by
\begin{eqnarray}
\mathbf{y}=\left(\underbrace{y_0,y_1}_{\text{subcarrier }1},\underbrace{y_2,y_3}_{\text{subcarrier }2},\cdots, \underbrace{y_{2N_s-2},y_{2N_s-1}}_{\text{subcarrier }N_s}\right).
\end{eqnarray}
Then, it is (modular 2) added to $\mathbf{z}$ to form two bits over each subcarrier. 

\begin{figure}[!t]
\centering
\includegraphics[width=0.35\textwidth]{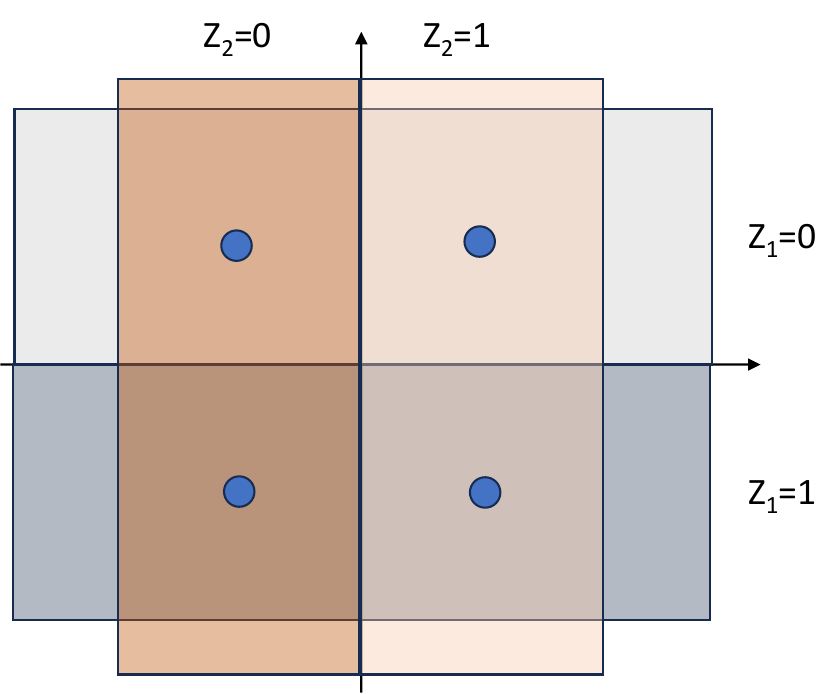}
\caption{Constellation partition used for high‑dimensional ($M=2$) shaping. MSB1 selects the inner/outer amplitude ring, MSB2 selects the sign, providing four power levels that the Viterbi metric exploits to equalize PSD.}
\label{fig:partition}
\end{figure}

The procedure until now is similar to the PAPR reduction, except for the goal function in \eqref{eq:ISL_metric}. However, the modulator that maps the bits $(\mathbf{y},\mathbf{b})$ to the point in the QAM constellation needs to be different. For PAPR, the MSB bits are used to partition the complex plane into quadrants, as illustrated in Fig. \ref{fig:partition}. In the complex plane, one bit in MSB identifies the side with respect to the vertical axis, while the other bit in MSB determines the side with respect to the horizontal axis. Therefore, the MSB bits can control the symbol phase, such that the symbols can better cancel each other in the sidelobes. In a contrast, in the context of waveform for ISAC, the goal is to control the power of each subcarrier in order to obtain a uniform PSD for the purpose of less sidelobes in the time domain. Therefore, the MSB bits need to distinguish different power allocations. Hence, in this paper, we propose the bit allocation to 16-QAM (suppose that 2 LSB bits are allocated to each subcarrier) illustrated in Fig. \ref{fig:QAM}, where 16 signal points are categorized into 4 subsets (marked by different colors), each corresponding to 2 MSB bits. The signals within each subset have the same power, regardless of the phase. Therefore, the 2 MSB bits control the power while the LSB bits determine the phase.  

\subsubsection{High-dimensional Constellation}
 An alternative approach is the high-dimensional constellation, namely using the symbols over $M_s$ subcarriers to form a high-dimensional symbol, where $M_s$ should divide $N_s$. The convolutional decoding output of $2N_s$ bits is given by
\begin{eqnarray}
\mathbf{y}=
\left(\underbrace{y_0,,...,y_{M_s-1}}_{\mathbf{y}_0},\cdots, \underbrace{y_{N_s-M_s},\cdots,y_{N_s-1}}_{\mathbf{y}_{\frac{N_s}{M_s}-1}}\right),
\end{eqnarray}
where each $y_k$ is a complex number over subcarrier $k$. Note that, when $M_s=1$, we attain the above sign-bit scheme. 

\begin{figure}[!t]
\centering
\includegraphics[width=0.35\textwidth]{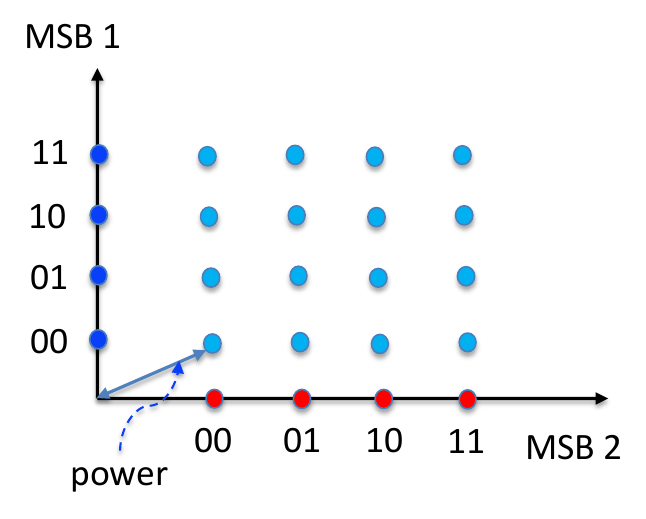}
\caption{2-dimensional power allocation for high dimensional signaling}
\label{fig:high_QAM}
\end{figure}

Then, the communication symbols are formed for every $M_s$ subcarriers. For simplicity, we assume that $M_s=2$, namely $y_0$ (2 bits) and $y_1$ (2 bits) form a 4-dimensional symbol (or a 2-complex-dimensional symbol). Similarly to the sign-bit signaling, we use the MSB bits to control the power. We assume that the 4 cases of 2 bits in subcarrier 1 form 4 levels of amplitudes over subcarrier 1 (see Fig.~\ref{fig:high_QAM}), and so is subcarrier 2. The amplitude levels are not necessarily the same for the two subcarriers. Then, for the composite 4-dimensional symbol, the corresponding power of the composite symbol is illustrated in Fig.~\ref{fig:high_QAM}. During the decoding procedure, determined by the Bellman equation \eqref{eq:Bellman}, the optimization is carried out for subcarriers 0 and 1 in a joint manner. The same procedure is repeated for the remaining $\frac{N_s}{M_s}-1$ subcarrier groups. 

\subsection{LSB Constellation}
In trellis shaping for PAPR reduction in OFDM signaling, the LSB bits allocation in the constellation has a significant impact on the performance, since the decoding procedure in the trellis shaping mainly optimizes the symbol phase, in order to cancel out the sidelobes in the autocorrelation. In the context of ISAC, since the goal is equivalent to minimizing the PSD variance, the phase has only a minor impact. Therefore, the LSB plays a much less important role in the trellis shaping in the ISL reduction in ISAC than in PAPR reduction in OFDM. According to our numerical results, random allocations of the LSB bits achieve similar performance levels in terms of sidelobes.  

\subsection{Communication-Sensing Trade-off}
Obviously, the performance of sensing is improved by the trellis shaping at the cost of lowering communication data rate. Take the sign-bit scheme for instance, the raw data for MSB is 1 bit per subcarrier and is extended to 2 bits by multiplying $\left(\mathbf{H}^{-1}\right)^T$. Meanwhile, 2 LSB bits are used for modulation directly. Therefore, each 16-QAM symbol encodes 3 bits, thus making the data rate $\frac{3}{4}$ of the original rate. As we will see, the substantial gain in the sensing performance justifies the drop in the communication data rate. 

\section{Joint Trellis Shaping for PAPR and ISL Reduction}\label{sec:joint}

In Section~\ref{sec:ISL_shaping}, we introduced the trellis-based waveform shaping framework aiming  to mitigate the time-domain sidelobes, typically measured by the ISL. In practice, however, a high PAPR can also be detrimental, as it may drive the power amplifier into its nonlinear region, causing signal distortion and degraded performance for both communication and sensing. This section extends the trellis shaping concept to simultaneously reduce time-domain sidelobes (thus low ISL) and PAPR. We formulate a combined cost function and integrate it into the same Viterbi (dynamic programming) framework.

\subsection{Fundamental Trade-off Between ISL and PAPR}

The design of radar waveforms involves optimizing multiple, often competing, performance metrics to meet system requirements. Among these objectives, the trade-off between ISL and PAPR stands out as a fundamental one, rooted in energy conservation principles and signal design constraints.

This ISL-PAPR trade-off arises from the conflicting requirements of sidelobe suppression versus amplitude uniformity. To minimize ISL, the waveform must exhibit an autocorrelation function (ACF) with suppressed sidelobes, which typically requires amplitude and/or phase modulation. However, such amplitude variations directly increase PAPR. Conversely, enforcing a constant-envelope (low-PAPR) waveform restricts modulation to phase-only, limiting the ability to suppress sidelobes. This duality can be understood through the Fourier transform relationship between the ACF and the PSD. For a waveform with PSD $S(f)$, the ACF is given by the inverse Fourier transform of the squared magnitude of the spectrum:
\begin{equation}
R[k] = \mathcal{F}^{-1}\{\, |S(f)|^2 \,\}\,,
\end{equation}
where $\mathcal{F}^{-1}\{\cdot\}$ denotes the inverse Fourier transform. To suppress ISL, one must shape $|S(f)|^2$ to approximate a rectangular spectrum (minimizing spectral leakage), which demands tapering the waveform's amplitude in the time domain. However, such tapering reduces the waveform's average power while preserving its peak amplitude, thereby increasing PAPR.

\begin{figure}[!t]
\centering
\includegraphics[width=0.45\textwidth]{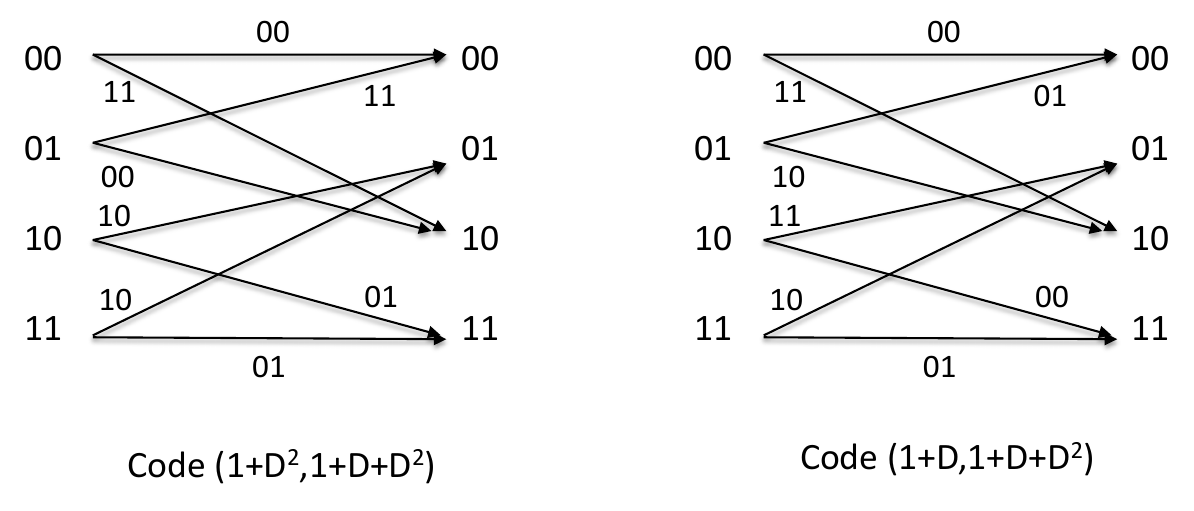}
\caption{State diagrams of the two rate‑½ convolutional codes tested: (a) $G=(1+D,1+D+D^2)$, (b) $G=(1+D²,1+D+D^2)$. Both have four states; empirical results show near‑identical shaping performance.}
\label{fig:trellis_conv}
\end{figure}

\subsection{Trellis Decoding with Dual Objectives}

To jointly optimize ISL and PAPR, we define a combined cost function that integrates both objectives into the trellis shaping process. Let $V_I(X)$ represent the ISL-related metric in \eqref{eq:ISL_metric}. For PAPR, We reuse the PAPR reduction metric in\cite{Ochiai2004} as the sidelobes of the frequency spectrum autocorrelation in \eqref{eq:autocorrelation_F}:
\begin{eqnarray}\label{eq:PAPR_metric}
V_{p}(X)=\sum_{k=0}^{N_s-l}X_k^*X_{k+l}.
\end{eqnarray}. 

The combined cost function is given by
\begin{equation}
  V(X; \alpha) = \alpha V_I(X) + (1-\alpha) V_p(X),    
\end{equation}
where $\alpha \in [0, 1]$ is a weighting factor that balances the trade-off between ISL and PAPR reduction. This weighted sum approach ensures that the two objectives are normalized within a single parameter range, thus facilitating a direct trade-off analysis. To ensure fair weighting, we normalize both metrics relative to their unshaped (baseline) values. We define the normalized costs as
\begin{equation}
    \widetilde{V}_I(X) = \frac{V_I(X)}{V_I^{\mathrm{unshaped}}}, \quad
\widetilde{V}_p(X) = \frac{V_p(X)}{V_p^{\mathrm{unshaped}}},
\end{equation}
where \( V_I^{\mathrm{unshaped}} \) and \( V_p^{\mathrm{unshaped}} \) are nominal reference values (e.g., those from an unshaped OFDM symbol). This ensures that both terms have comparable magnitudes when entering the Viterbi metric, namely
\begin{equation}
    V^{\mathrm{norm}}(X; \alpha) = \alpha \widetilde{V}_I(X) + (1 - \alpha) \widetilde{V}_p(X).
\end{equation}

The trellis shaping process follows that in Section~\ref{sec:ISL_shaping}, whereas now evaluating the state transitions using the combined metric. For trellis state $ \phi_t $ at subcarrier $ t $, the recursive cost is given by
\begin{equation}
    V(\phi_t) = \min_{\phi_{t-1} \rightarrow \phi_t} \left[ V(\phi_{t-1}) + \Delta (\phi_{t-1}, \phi_t; \alpha) \right], 
\end{equation}
with the boundary condition \( V(\phi_{N_s+1}) = 0 \). The incremental cost is given by
\begin{equation}
    \Delta(\phi_{t-1}, \phi_t; \alpha) = \alpha \Delta_I^{\mathrm{norm}}(\phi_{t-1}, \phi_t) + (1 - \alpha) \Delta_p^{\mathrm{norm}}(\phi_{t-1}, \phi_t),
\end{equation}
where the quantities are defined as
\begin{equation}
    \begin{aligned}
        \Delta_I^{\mathrm{norm}}(\phi_{t-1}, \phi_t) = \frac{\Delta_I(\phi_{t-1}, \phi_t)}{V_I^{\mathrm{unshaped}}} , \\
\Delta_p^{\mathrm{norm}}(\phi_{t-1}, \phi_t) = \frac{\Delta_p(\phi_{t-1}, \phi_t)}{V_p^{\mathrm{unshaped}}}, 
    \end{aligned}
\end{equation}
are the normalized incremental contributions to ISL and PAPR, respectively. The ISL incremental cost \( \Delta_I(\phi_{t-1}, \phi_t) \) follows the definitions in \eqref{eq:ISL_metric}. The PAPR-related metric for trellis state $\phi_t$ evolves recursively as
\begin{equation}
    V_p(\phi_t) = \min_{\phi_{t-1} \rightarrow \phi_t} \left[ V_p(\phi_{t-1}) + \Delta_p(\phi_{t-1}, \phi_t) \right],
\end{equation}
where the incremental PAPR cost is derived in\cite{Ochiai2004} as
\begin{equation}
\Delta_p(\phi_{t-1}, \phi_t) = \sum_{m=1}^{t-2} 2 \Re \left\{ R_m^{(t-1)*} \delta_m^{(t-1)} \right\}
+ \sum_{m=1}^{t-1} \left| \delta_m^{(t-1)} \right|^2.
\end{equation}

Here, $R_m$ is the aperiodic autocorrelation function of the complex OFDM data sequence and $\delta^{(i)}_m \triangleq X_i X_{i - m}^*$. The Viterbi algorithm selects the MSB bits $ \mathbf{y} $ at each step to minimize $ V(\phi_t) $, shaping the waveform to balance the PSD uniformity (low ISL) and the phase alignment (low PAPR). 

\begin{figure}[!t]
 \centering
 \includegraphics[width=2.8in]{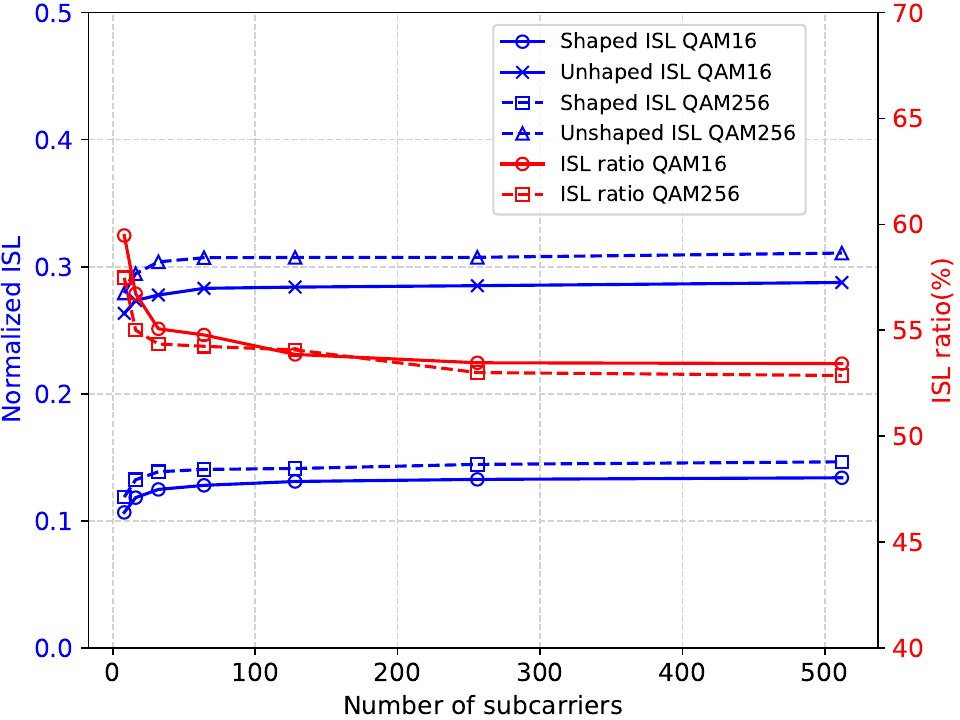}\\
\caption{Comparison of ISL using 16QAM and 256QAM, across varying numbers of subcarriers.}
 \label{fig:ISL_QAM16_256}
\end{figure}

\begin{figure}[!t]
 \centering
 \includegraphics[width=2.8in]{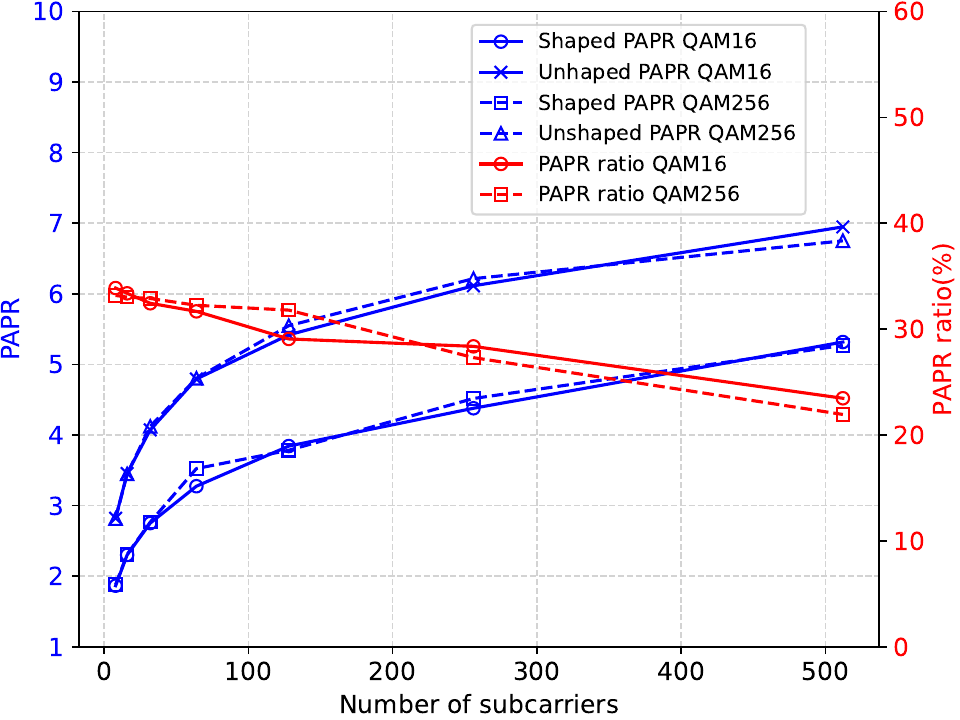}\\
\caption{Comparison of PAPR using 16QAM and 256QAM, across varying numbers of subcarriers.}
 \label{fig:PAPR_QAM16_256}
\end{figure}

\subsection{Normalization and Stability}

When $N_s$ is large, the ISL metric $ V_I = V_{\text{small}}$ is discrete, reflecting finite power levels, while $ V_p $ is continuous, with larger variance due to the cumulative autocorrelation terms. A key challenge arises from balancing the two cost metrics $ V_I $ and $V_p$ when their scales and variances differ significantly. If one cost function varies much more sharply than the other, it can dominate the optimization and lead to several issues:
\begin{itemize}\setlength\itemsep{0em}
    \item The Viterbi algorithm, which greedily minimizes the combined cost at each step, may overreact to minor fluctuations in the rapidly varying cost. This can result in “jittery” or unstable paths that are locally optimal but globally suboptimal.
    \item Early decisions dominated by the more sensitive cost metric may become effectively irreversible, even if they later prove detrimental to the overall solution.
    \item If one cost varies much faster than the other, small changes in the fast-changing metric can disproportionately influence the total cost. The decoder may become overly sensitive to the rapidly varying objective while neglecting the other.
\end{itemize}

To mitigate these issues, we propose a normalization strategy when $N_s$ is large. Instead of applying a direct constraint to PAPR, we discretize its continuous cost by introducing a threshold $\epsilon$ to mitigate the sensitivity issues and align it more closely with the discrete nature of the ISL metric. The modified PAPR-related cost is thus defined as

\begin{figure}[!t]
 \centering
 \includegraphics[width=2.8in]{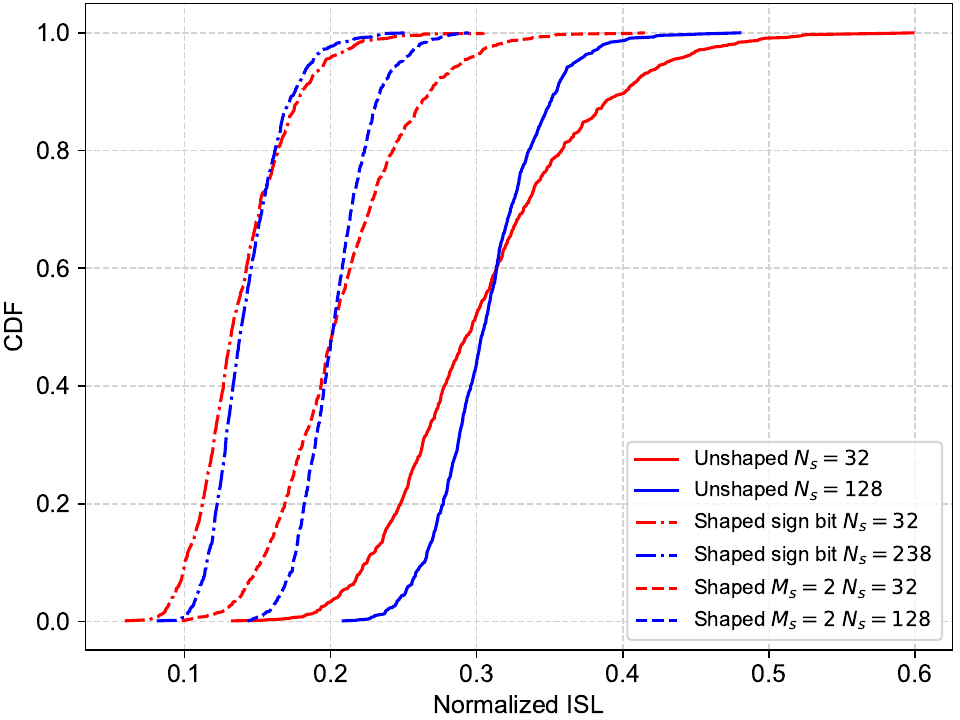}\\
\caption{CDFs of ISL for multidimensional trellis shaping.}
 \label{fig:multidimension_ISL}
\end{figure}

\begin{figure}[!t]
 \centering
 \includegraphics[width=2.8in]{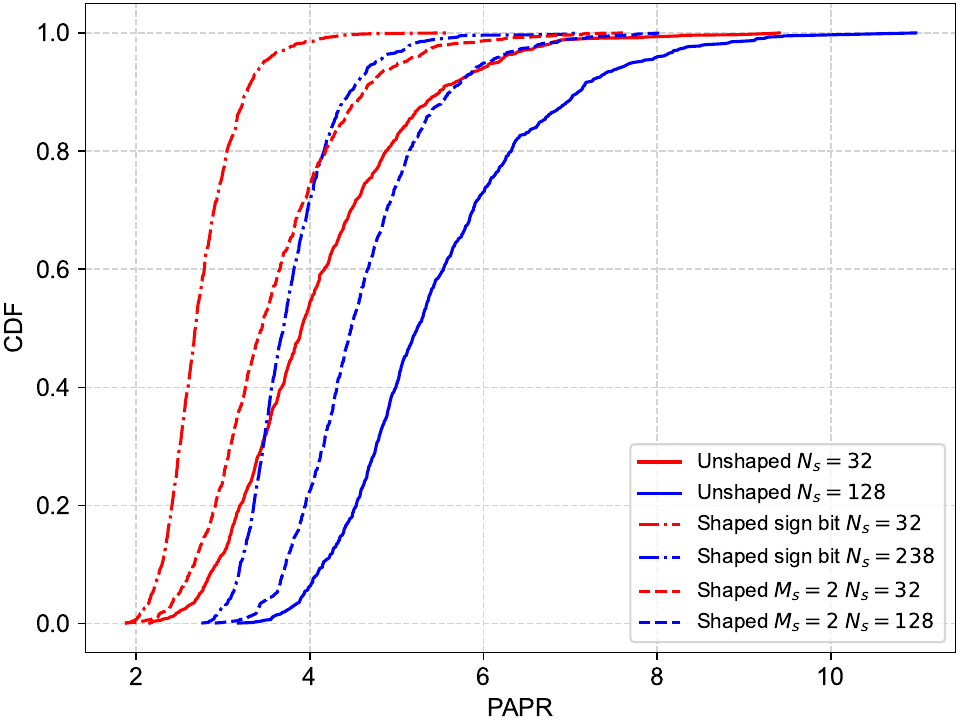}\\
\caption{CDFs of PAPR for multidimensional trellis shaping.}
 \label{fig:multidimension_PAPR}
\end{figure}

\begin{equation}
V_p(X_k) = \begin{cases}
\frac{P_t}{N_s}, & \text{if } R(k) > \epsilon, \\
-\frac{P_t}{N_s}, & \text{if } R(k) \leq \epsilon,
\end{cases}
\end{equation}
where $R(k)$ is the instantaneous PAPR-related cost at the $k$-th step. If this cost exceeds the threshold $\epsilon$, a positive penalty is added, thus discouraging high-PAPR solutions. Conversely, if the cost is below the threshold, a negative reward encourages low-PAPR solutions. This approach reduces abrupt fluctuations in the PAPR metric, making it comparable in the sensitivity to the ISL metric.

Note that the threshold $\epsilon$ is tunable and depends on the number of subcarriers. Specifically, as $N_s$ increases, the amplitude distribution of OFDM signals approaches a Gaussian distribution due to the Central Limit Theorem, inherently increasing the PAPR. Consequently, a larger threshold $\epsilon$ is typically required for systems with more subcarriers. Empirical results suggest $\epsilon = 3.5$ as a practical initial setting, with subsequent adjustments based on specific system characteristics and requirements.

\section{Numerical and Experimental Results}\label{sec:numerical}



In this section, we use numerical simulation results to demonstrate the performance of the proposed trellis-shaping-based waveform synthesis for ISAC.

\subsection{Convolutional Codes}
We adopt the convolutional code with generating matrix $(1+D^2,1+D+D^2)$ of data rate $\frac{1}{2}$. For comparison, we also tested another code with generating matrix $(1+D,1+D+D^2)$. The trellis of each code is shown in Fig. \ref{fig:trellis_conv}. We observe that the trellises of these two codes are similar.

\begin{figure}[!t]
\centering
\includegraphics[width=0.4\textwidth]{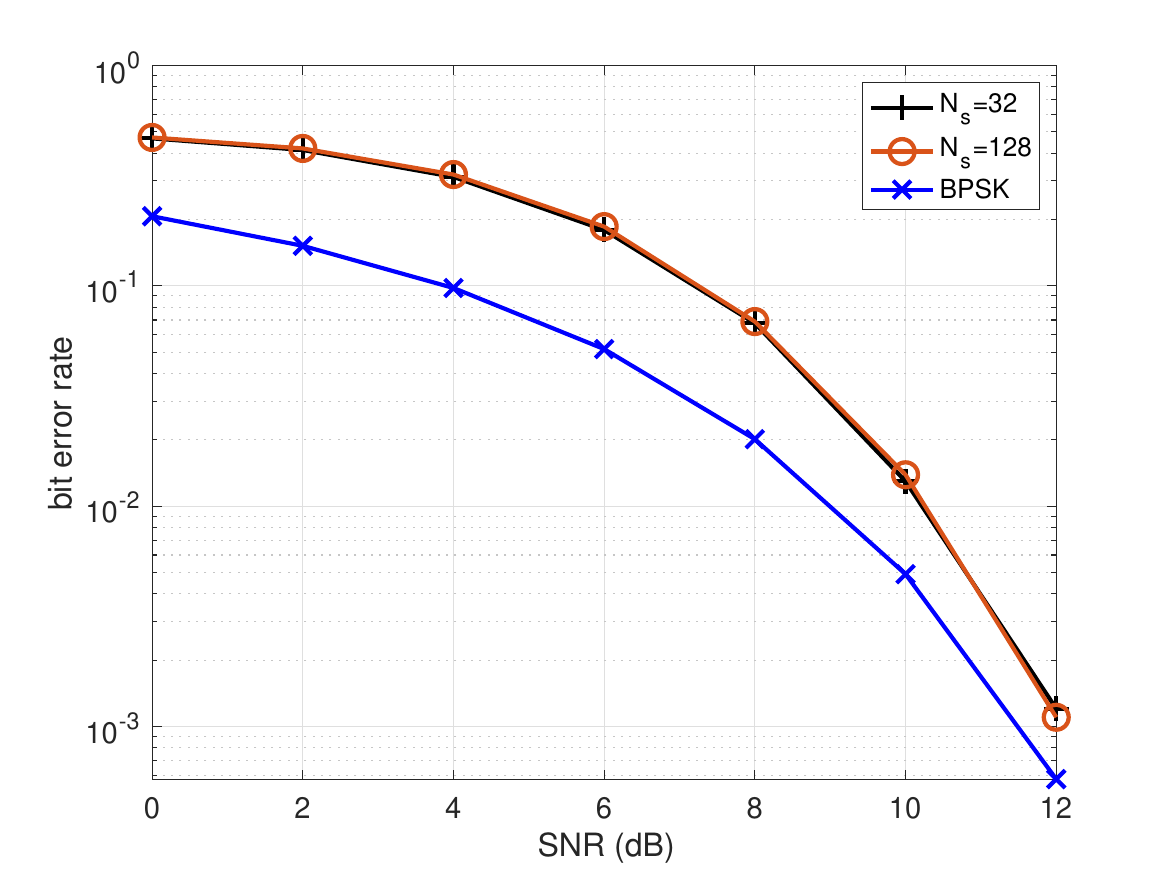}
\caption{Bit error rate of MSB in trellis shaping}
\label{fig:BER}
\end{figure}

\subsection{ISL and PAPR Reductions}

We evaluated the trellis shaping performance for ISL and PAPR separately using different constellation sizes, 16QAM and 256QAM. The sign-bit scheme was employed to obtain the results. We define two metrics to quantify the ISL and PAPR reduction performance:
\begin{equation}
\begin{aligned}
    \text{ISL}_{\text{ratio}} &= \frac{ \text{ISL}_{\text{unshaped}} - \text{ISL}_{\text{shaped}}  }{\text{ISL}_{\text{unshaped}}}, \\
    \text{PAPR}_{\text{ratio}} &= \frac{ \text{PAPR}_{\text{unshaped}} - \text{PAPR}_{\text{shaped}}  }{\text{PAPR}_{\text{unshaped}}}.
\end{aligned}  
\end{equation}

We first assess the optimization performance of ISL by exclusively optimizing ISL itself (with $\alpha = 1$). Fig.~\ref{fig:ISL_QAM16_256} presents the ISL values for both shaped and unshaped waveforms across various numbers of subcarriers and constellation sizes. It can be observed that as the number of subcarriers increases, the ISL initially rises slightly and then remains nearly constant. This behavior occurs because a larger number of subcarriers tends to induce greater variance in the PSD, consequently degrading the ISL. Additionally, the ISL of 256QAM is consistently higher than that of 16QAM due to 256QAM possessing a greater number of power levels, resulting in a less uniform PSD and inferior sidelobe performance. Furthermore, the ISL improvement ratio diminishes as the number of subcarriers increases. Notably, the ISL improvement ratios for 16QAM and 256QAM are almost identical.

We then evaluate the optimization performance of PAPR when only ISL is optimized (with $\alpha = 0$). Fig.~\ref{fig:PAPR_QAM16_256} illustrates the PAPR values for shaped and unshaped waveforms with varying constellation sizes and numbers of subcarriers. The results show that increasing the number of subcarriers leads to a corresponding increase in PAPR. Similar to the ISL results, the PAPR improvement ratio declines with an increased number of subcarriers, and the PAPR improvement ratios for 16QAM and 256QAM remain nearly identical.

\subsection{Multidimensional}

The application of multidimensional trellis shaping effectively reduces both the redundancy ratio for a given constellation size and the complexity of the shaping Viterbi decoder. However, this cost reduction is accompanied by a relative degradation in performance, specifically in terms of dynamic range reduction capability.

The numerical results utilize 256QAM modulation with two subcarrier configurations: $N_s = 32$ and $128$. Comparisons of the performances of the two shaping distributions in terms of ISL and PAPR reduction for dimensionalities of 1 (sign-bit) and 2 are illustrated separately in Figures~\ref{fig:multidimension_ISL} and \ref{fig:multidimension_PAPR}. As shown in Fig.~\ref{fig:multidimension_ISL}, increasing dimensionality from sign-bit to 2 enhances ISL performance. A similar trend, depicted in Fig.~\ref{fig:multidimension_PAPR}, indicates that higher dimensionality results in diminished dynamic range reduction performance.

\begin{figure}[!t]
 \centering
 \includegraphics[width=2.8in]{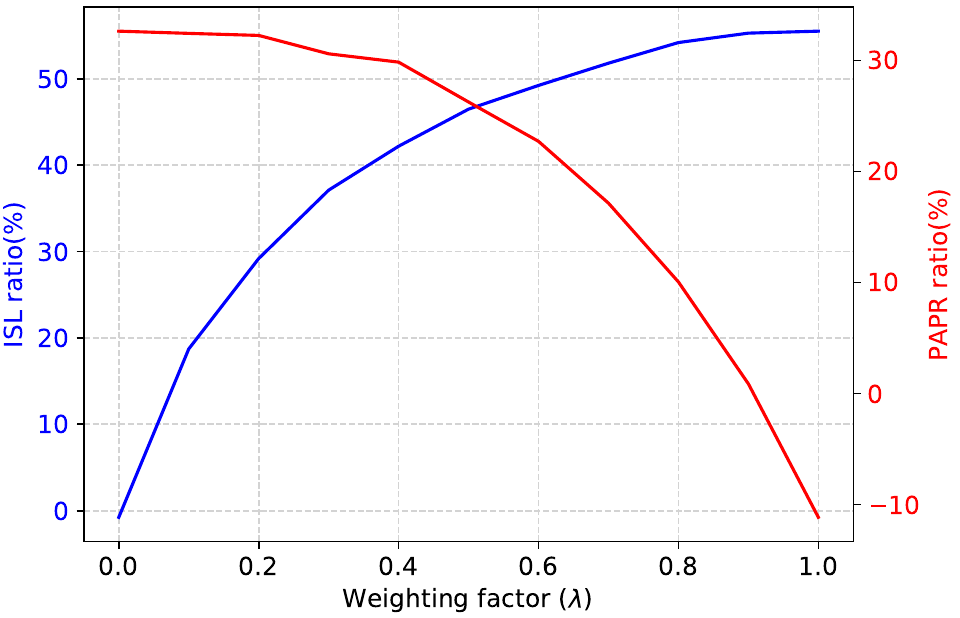}\\
\caption{ISL‑ and PAPR‑improvement ratios versus weighting factor $\lambda$ ($N_s=32$, 16QAM)}
 \label{fig:ISL_PAPR}
\end{figure}

\begin{figure*}[!t]
    \centering
    \subfigure[Power of each subcarrier.]
    {
        \label{fig:sim1a}
        \includegraphics[width=1.95in]{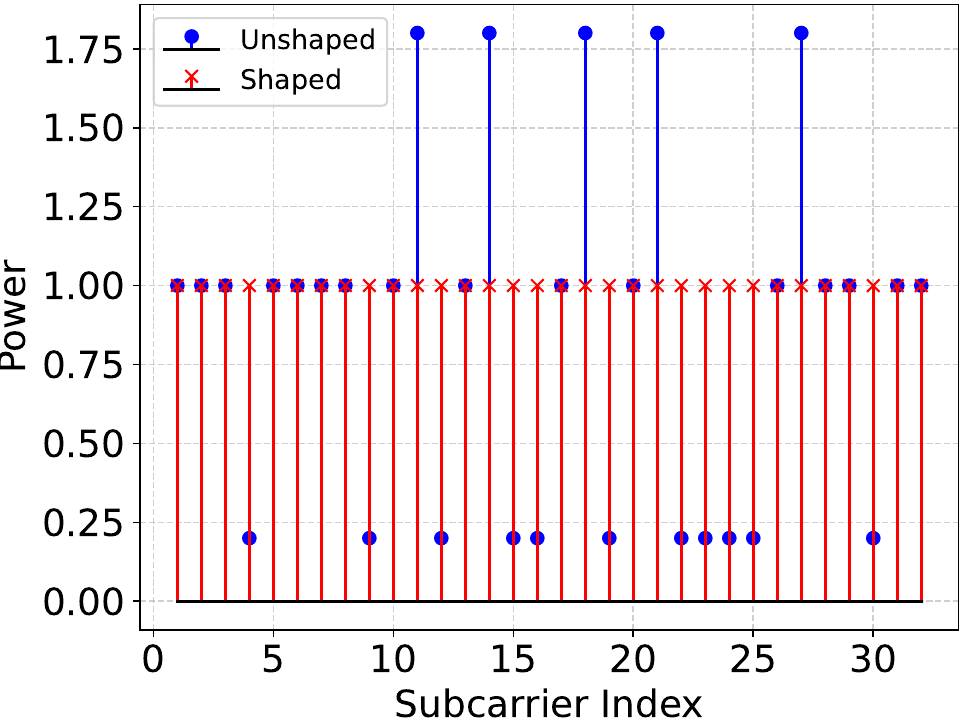}
    }
    \hspace{0.01\linewidth}
    \subfigure[Normalized auto correlations.]
    {
        \label{fig:sim1b}
        \includegraphics[width=1.95in]{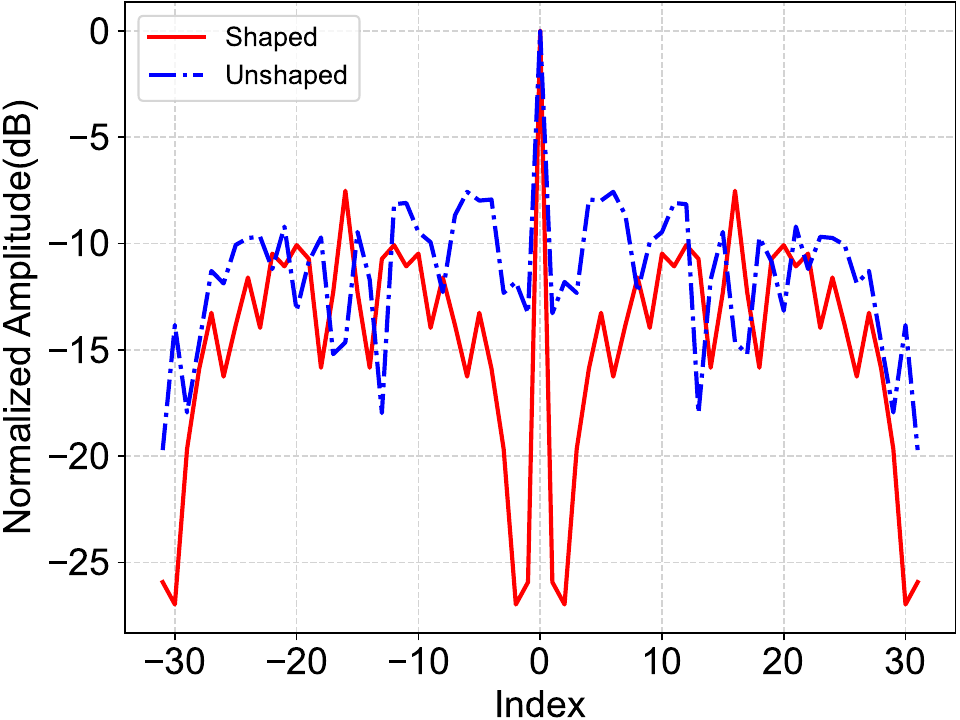}
    }
    \hspace{0.01\linewidth}
    \subfigure[Power profile in time domain.]
    {
        \label{fig:sim1c}
        \includegraphics[width=1.95in]{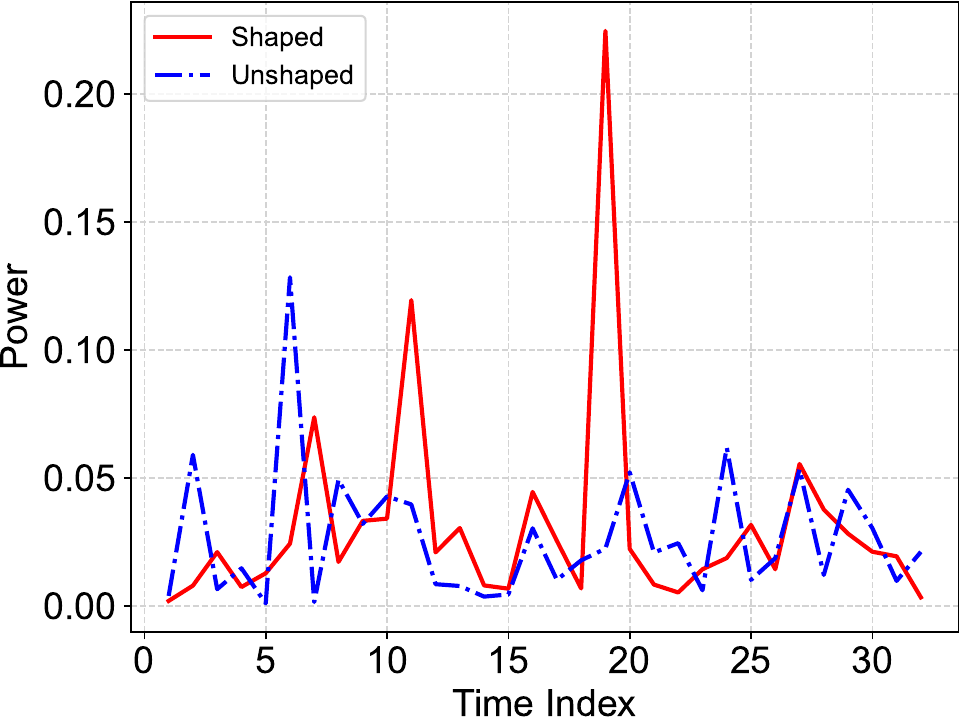}
    }
    \hspace{0.01\linewidth}
\caption{Simulation results with $\lambda=1$ (optimizaing ISL).}
\label{fig:sim1}
\end{figure*}

\begin{figure*}[!t]
    \centering
    \subfigure[Power of each subcarrier.]
    {
        \label{fig:sim2a}
        \includegraphics[width=1.95in]{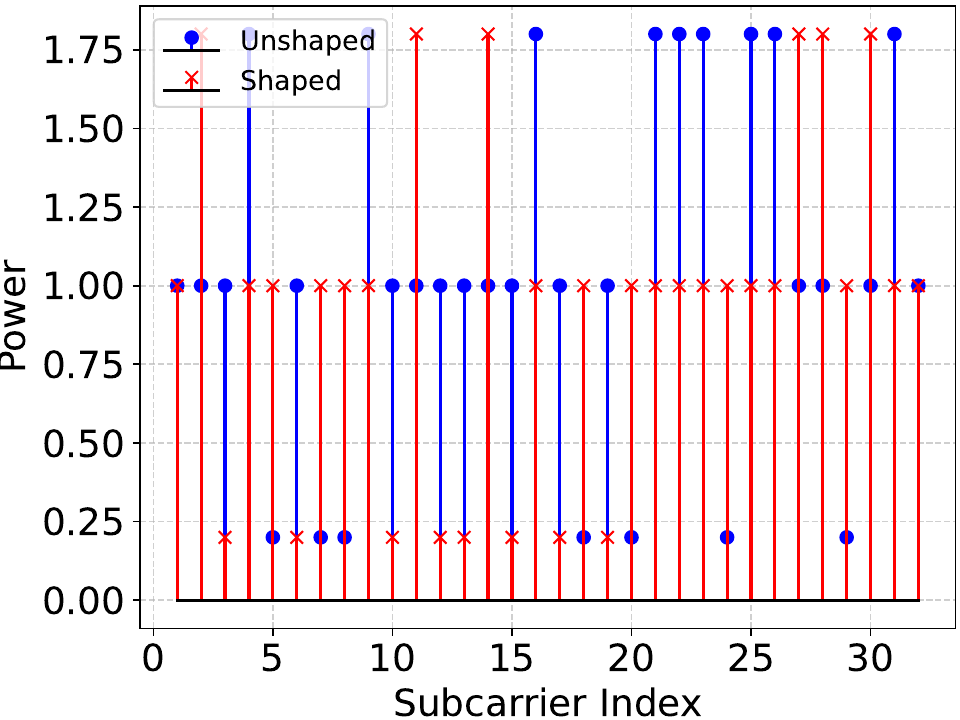}
    }
    \hspace{0.01\linewidth}
    \subfigure[Normalized auto correlations.]
    {
        \label{fig:sim2b}
        \includegraphics[width=1.95in]{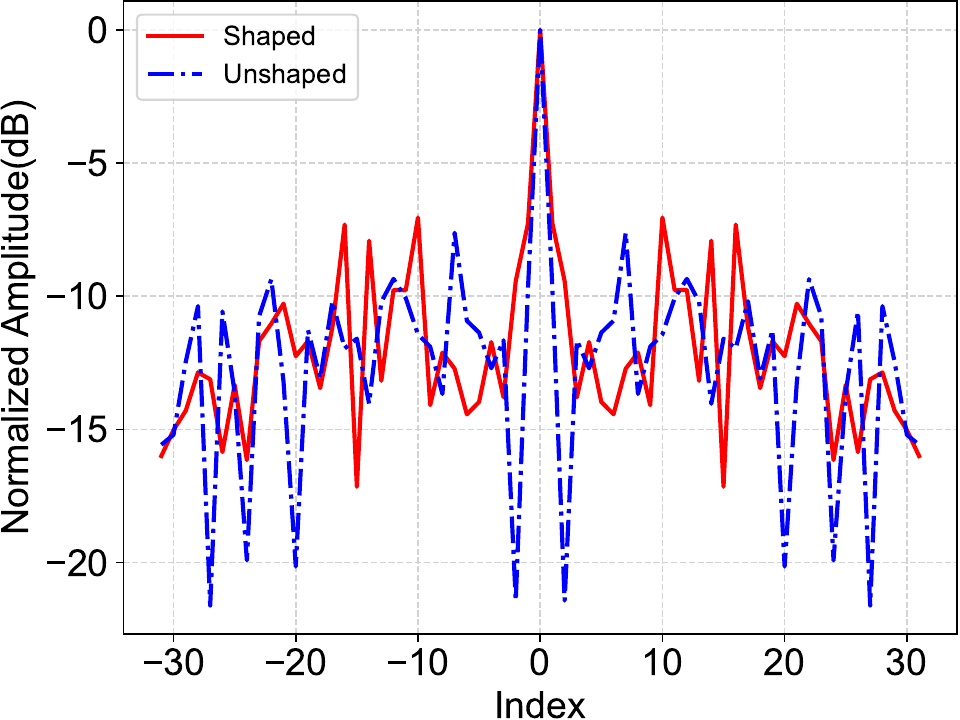}
    }
    \hspace{0.01\linewidth}
    \subfigure[Power profile in time domain.]
    {
        \label{fig:sim2c}
        \includegraphics[width=1.95in]{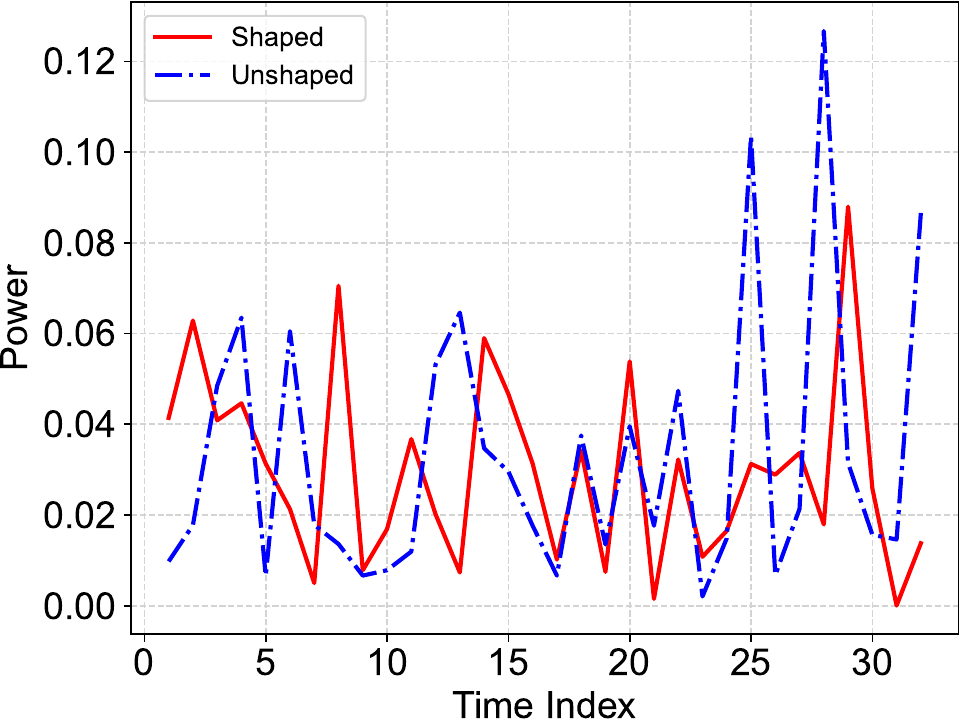}
    }
    \hspace{0.01\linewidth}
\caption{Simulation results with $\lambda=0$ (optimizaing PAPR).}
\label{fig:sim2}
\end{figure*}

\begin{figure*}[!t]
    \centering
    \subfigure[Power of each subcarrier.]
    {
        \label{fig:sim3a}
        \includegraphics[width=1.95in]{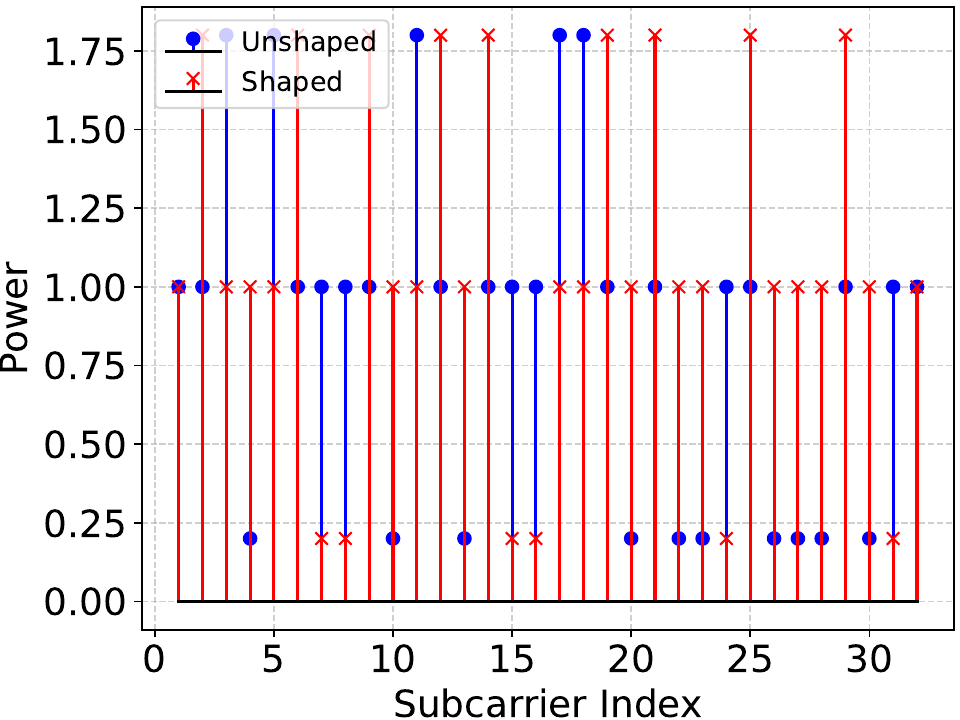}
    }
    \hspace{0.01\linewidth}
    \subfigure[Normalized auto correlations.]
    {
        \label{fig:sim3b}
        \includegraphics[width=1.95in]{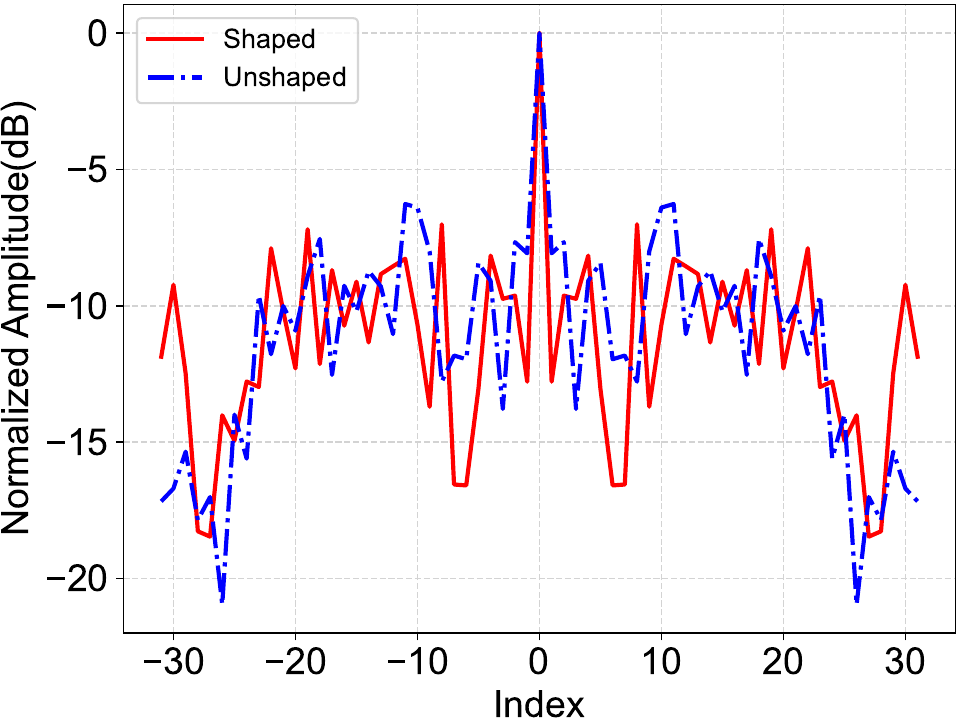}
    }
    \hspace{0.01\linewidth}
    \subfigure[Power profile in time domain.]
    {
        \label{fig:sim3c}
        \includegraphics[width=1.95in]{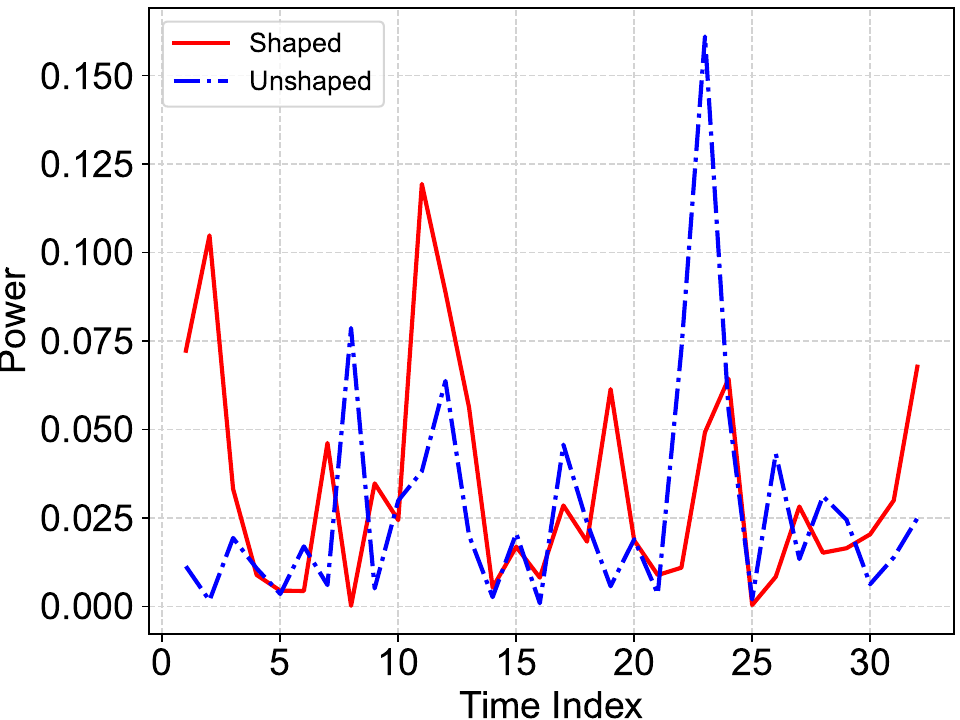}
    }
    \hspace{0.01\linewidth}
\caption{Simulation results with $\lambda=0.5$ (joint optimizaing PAPR and ISL).}
\label{fig:sim3}
\end{figure*}

\subsection{Communication Performance}
The original bits in $\mathbf{s}$ can be recovered by removing $\mathbf{y}$ with $\left(\mathbf{H}^{-1}\right)^T$. However, this depends on the assumption that the demodulation is free of errors. In noisy channels, when there exist errors, it is not clear whether $\mathbf{s}$ can still be recovered (despite some errors) as if $\mathbf{y}$ does not exist. Therefore, we tested the performance of bit error rate at the receiver, by assuming Gaussian noise with different signal-to-noise ratios (SNRs). 
We use the code with $G=(1+D^2,1+D+D^2)$. The corresponding parity check matrix is given by $H^T=(1+D+D^2,1+D^2)^T$, with the generalized inverse $\left(H^{-1}\right)^T=(D,1+D)$. The bit error rates of cases $N_s=32$ and $N_s=128$ are calculated for different values of SNR. We also calculated the bit error rate of Binary Phase Shift Keying (BPSK) for comparison. The results are plotted in Fig. \ref{fig:BER}. We observe that there is a significant gap between the bit error rates (BERs) of the trellis shaping and BPSK. Therefore, the increased BER also reduces the channel capacity, in addition to the lower data rate. 

\subsection{PAPR-ISL Tradeoff}

We evaluate the trade-off performance of the joint optimization of PAPR and ISL. The numerical results are obtained using 16QAM modulation with a subcarrier number of $N_s = 32$.
As illustrated in Fig.~\ref{fig:ISL_PAPR}, the weighting factor $\lambda$ controls the trade-off between ISL and PAPR optimizations. When $\lambda = 0$, the trellis shaping is solely focused on minimizing ISL, achieving an ISL improvement of over 30\%, while PAPR performance deteriorates, as indicated by a negative PAPR ratio. Conversely, when $\lambda = 1$, the system prioritizes the PAPR optimization, yielding a PAPR reduction of over 30\% at the expense of ISL degradation. Furthermore, as $\lambda$ varies from 0 to 1, the ISL ratio consistently decreases, while the PAPR ratio increases, demonstrating a clear trade-off between these two metrics. This result empirically validates the negative correlation between ISL and PAPR.

To further illustrate the effectiveness of joint trellis shaping, we visualize the power allocation across subcarriers, the normalized autocorrelation results, and the signal power in the time domain. Fig.~\ref{fig:sim1} presents the results for the case where $\lambda = 1$, focusing solely on ISL optimization without considering PAPR effects. In Fig.~\ref{fig:sim1a}, it can be observed that the optimal solution results in an approximately uniform power allocation across subcarriers to mitigate sidelobes, aligning with our previous analysis. Fig.~\ref{fig:sim1b} shows the normalized autocorrelation results for both shaped and unshaped signals. The shaped signal exhibits significantly lower sidelobe power compared to the unshaped signal, demonstrating the effectiveness of ISL mitigation, which ultimately enhances sensing performance. In Fig.~\ref{fig:sim1c}, the power profile of the signal in the time domain is presented to visualize the PAPR characteristics. It is evident that the shaped signal contains sharper peaks with higher power, leading to a degradation in the PAPR performance. Fig.~\ref{fig:sim2} depicts the results for $\lambda = 0$, where PAPR is optimized without considering ISL. In Fig.~\ref{fig:sim2a}, unlike the uniform power allocation observed in Fig.~\ref{fig:sim1}, the power distribution does not remain constant. This is because optimizing PAPR requires careful joint design of both the phase and amplitude of each subcarrier. In Fig.~\ref{fig:sim2b}, the shaped signal exhibits an increased autocorrelation sidelobe level compared to the ISL-focused case, while Fig.~\ref{fig:sim2c} clearly demonstrates a significant PAPR reduction. These results further validate the inherent trade-off between ISL and PAPR. Finally, in Fig.~\ref{fig:sim3}, we illustrate the effectiveness of the joint optimization scheme by selecting $\lambda = 0.5$, where both ISL and PAPR are considered. As shown in Fig.~\ref{fig:sim3b} and Fig.~\ref{fig:sim3c}, the performances of both ISL and PAPR are improved. Although the improvements are less pronounced compared to optimizing a single metric, this result reinforces the fundamental trade-off between ISL and PAPR. Moreover, it highlights the flexibility of adjusting the optimization focus based on different application requirements.

\subsection{Experimental Results}

To substantiate the efficacy of the proposed trellis waveform shaping method, we conducted comprehensive experiments using a software-defined radio testbed operating in the mmWave band. These experiments serve as a critical validation, demonstrating that the method performs robustly not only in simulations but also in practical, real-world conditions.

\begin{figure}[!t]
   \centering
   \subfigure[Communication TX and monostatic radar.]
   {
       \includegraphics[width=0.45\columnwidth]{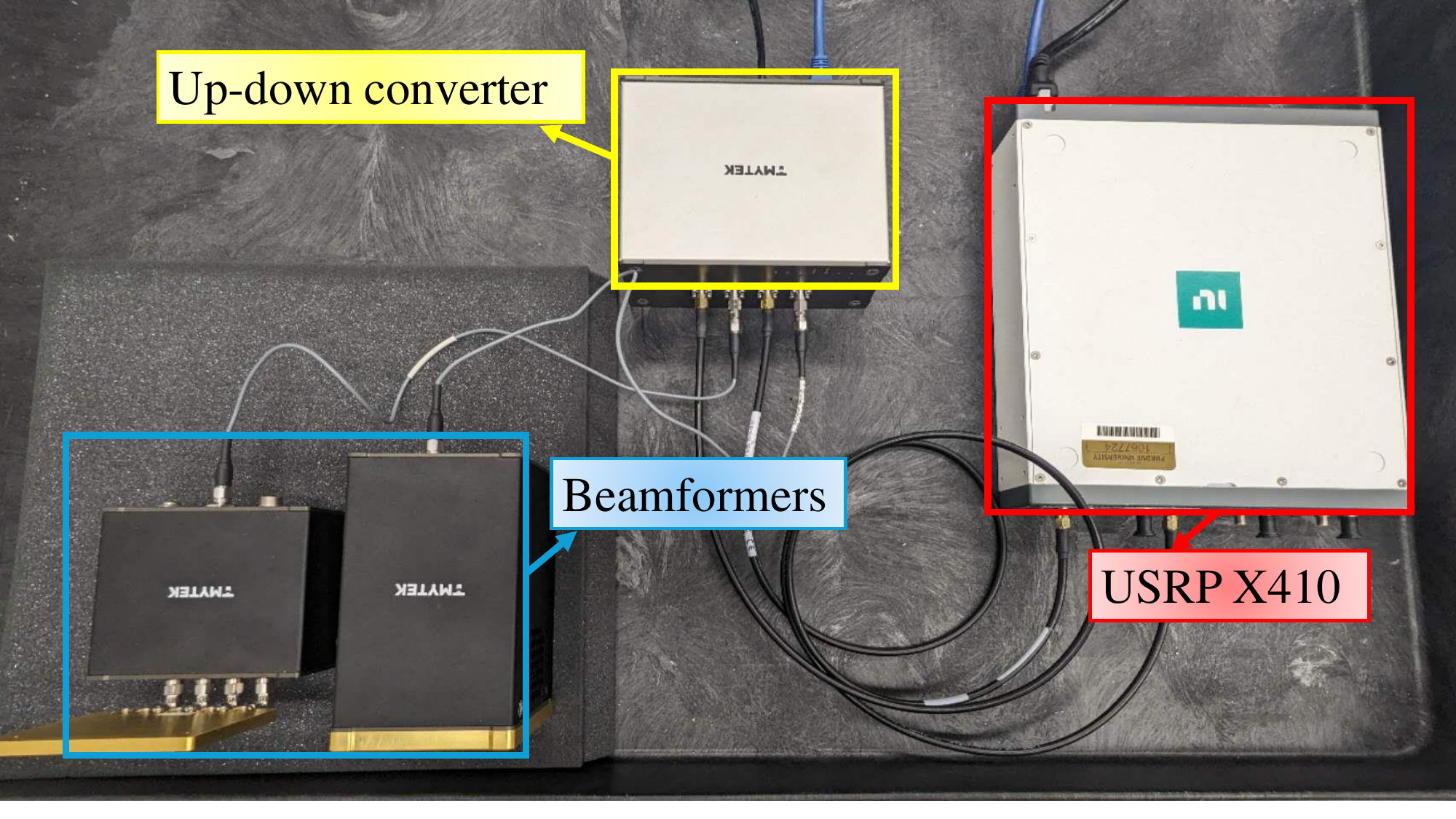}
       \label{fig:TX}
   }
   \subfigure[Communication RX.]
   {
      \includegraphics[width=0.45\columnwidth]{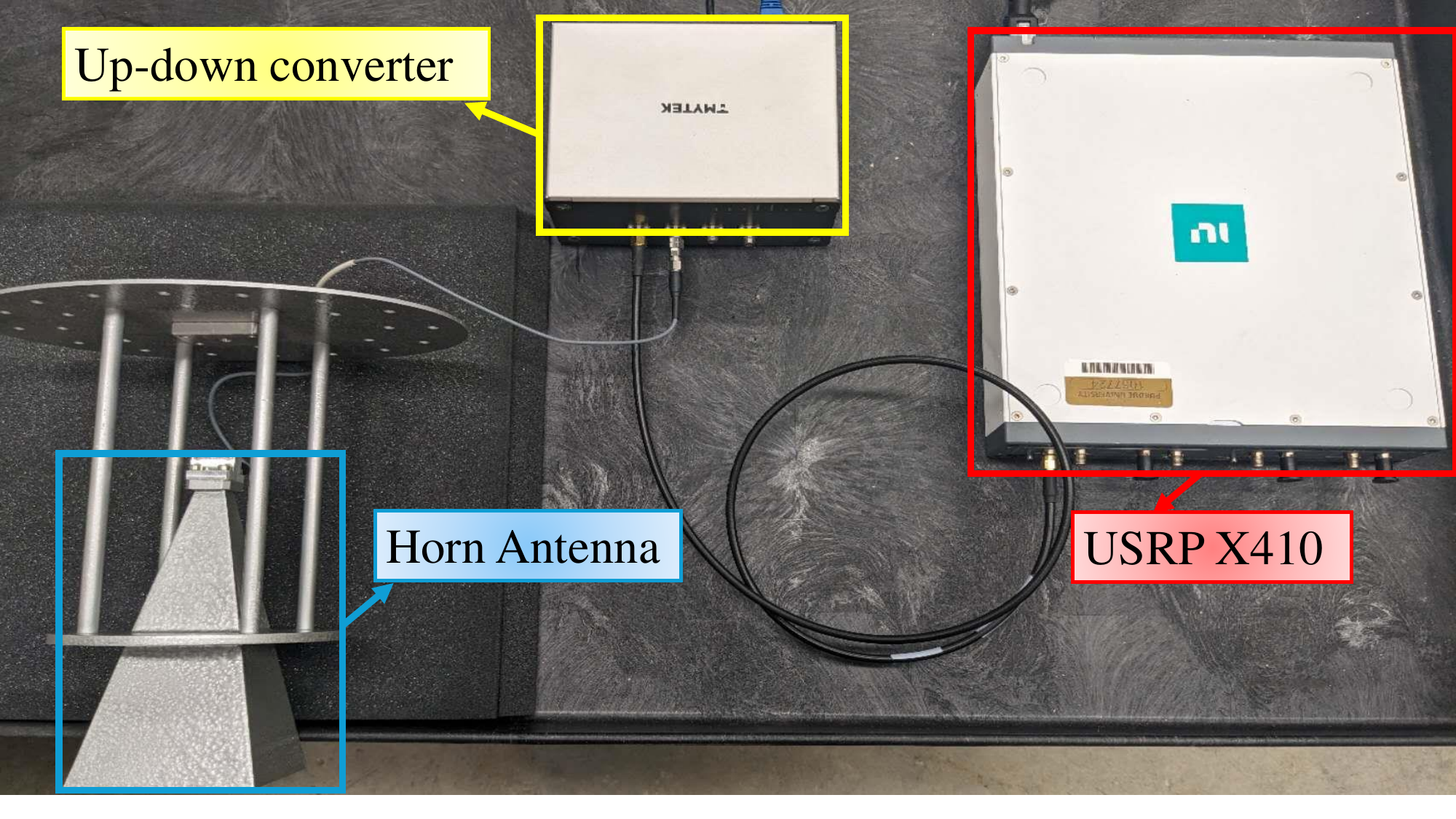}
      \label{fig:RX}
   }
   \subfigure[Radar estimation setup.]
   {
       \includegraphics[width=0.45\columnwidth]{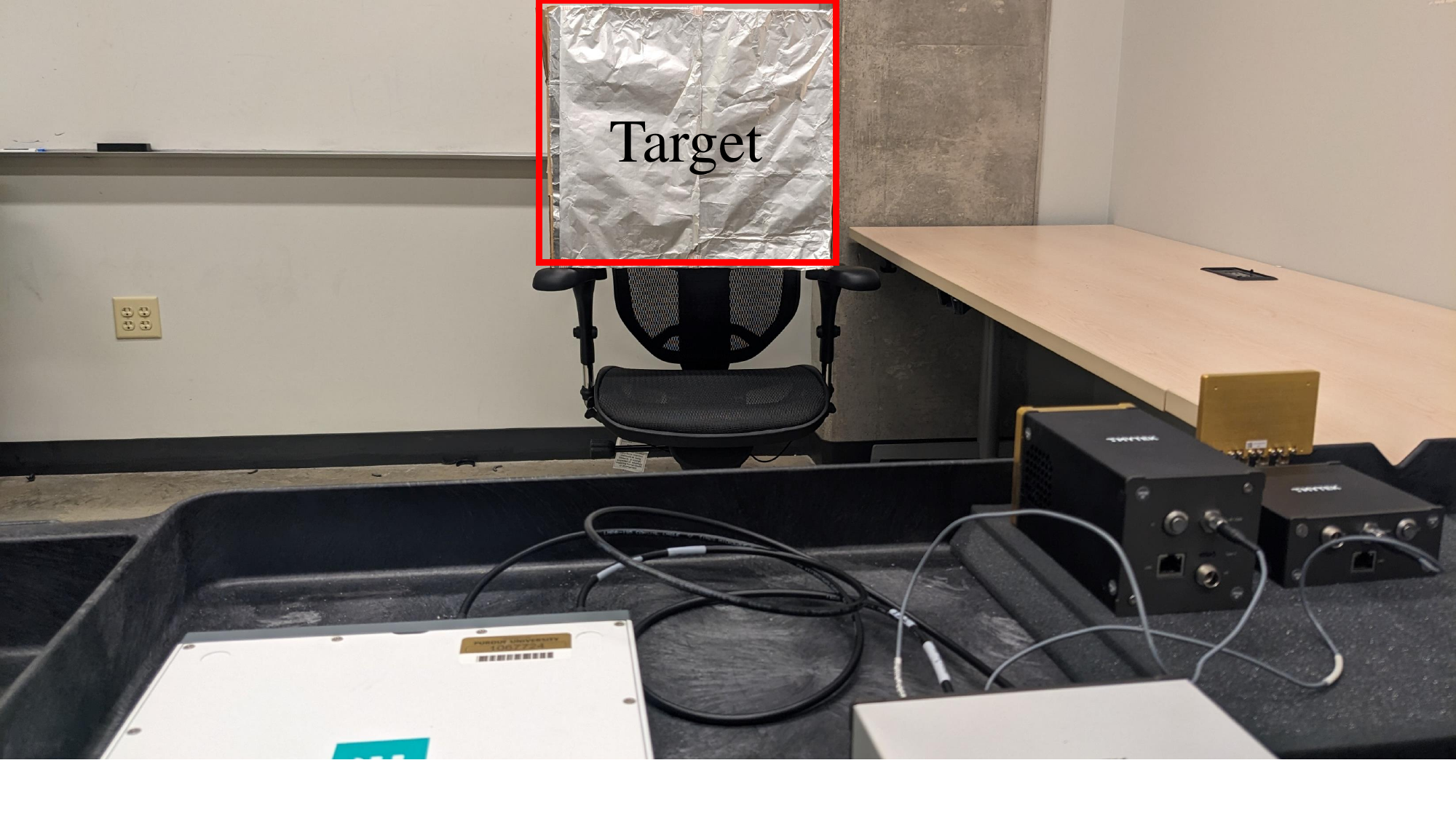}
       \label{fig:experiment_setting_radar}
   }
   \subfigure[Communication setup.]
   {
      \includegraphics[width=0.45\columnwidth]{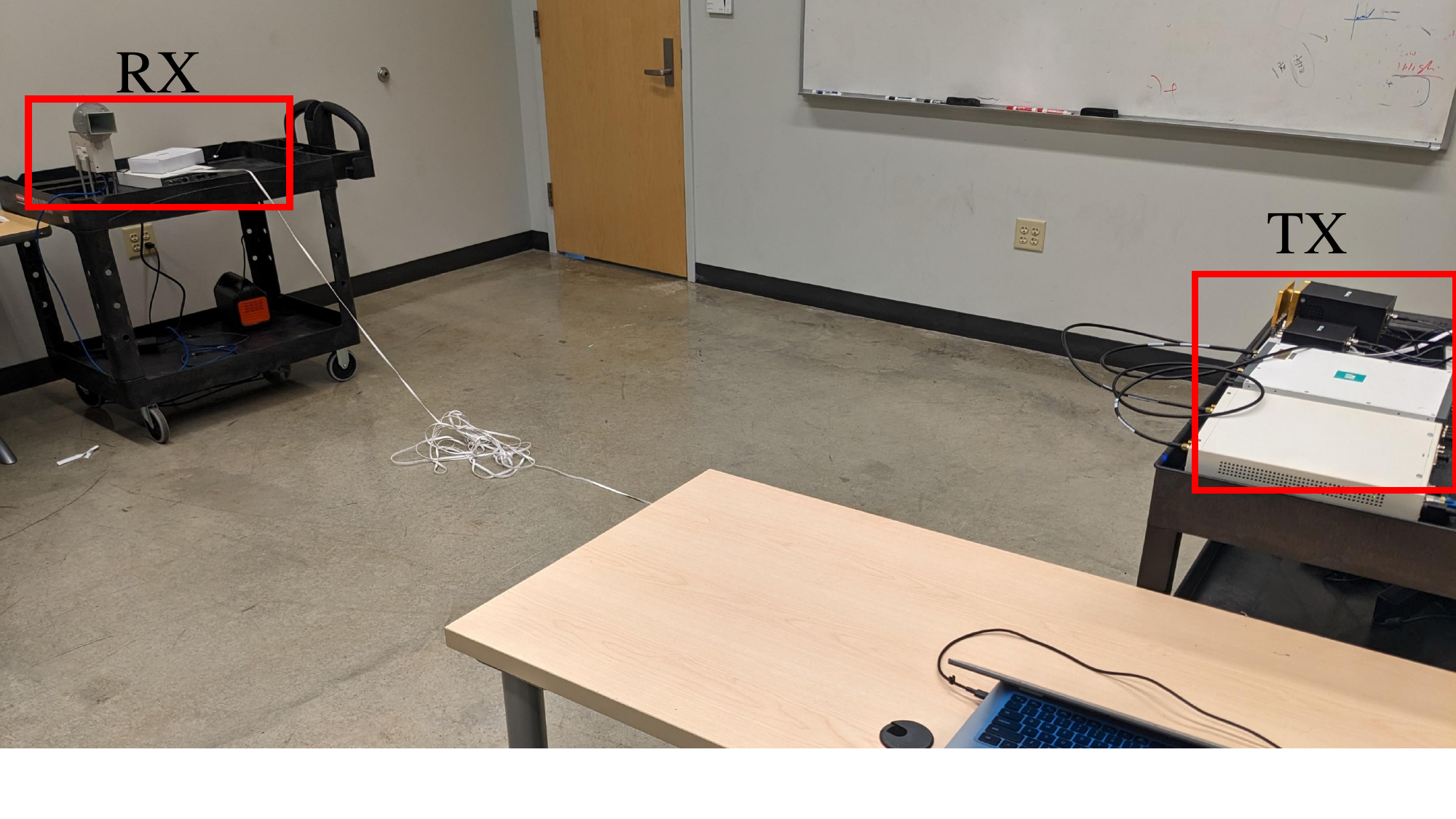}
      \label{fig:experiment_setting_comms}
   }
   \caption{Hardware and experimental setup.}
   \label{fig:experiment_setting}
\end{figure}

\begin{figure}[!t]
   \centering
   \subfigure[ISL.]
   {
       \label{fig:exp1}
        \includegraphics[width=0.43\columnwidth]{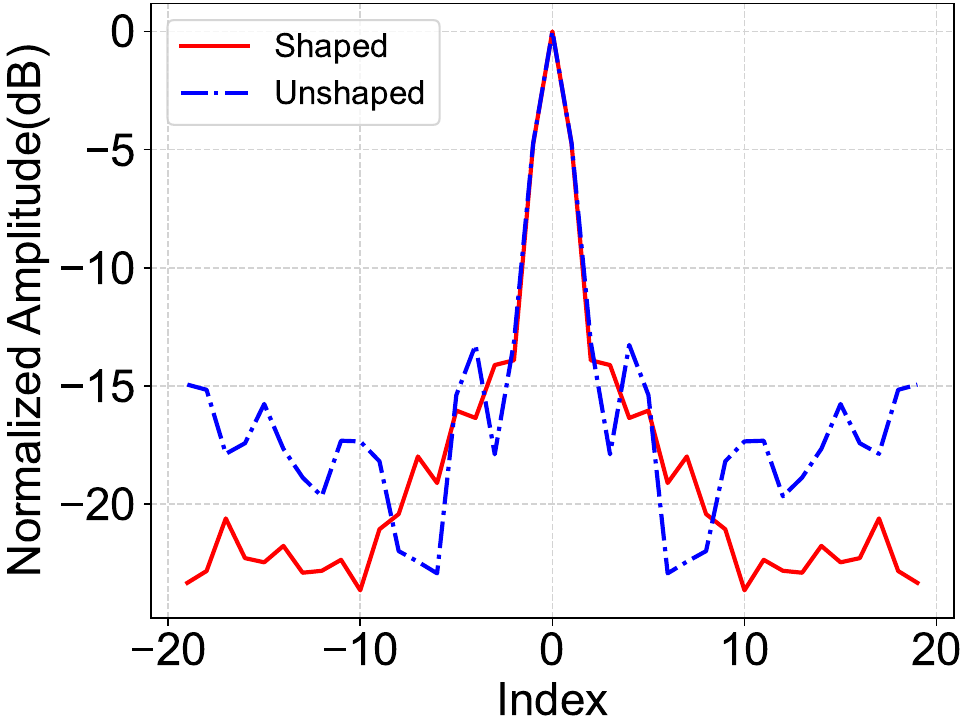}
   }
   \hspace{0.001\linewidth}
   \subfigure[PAPR.]
   {
      \label{fig:exp2}
        \includegraphics[width=0.43\columnwidth]{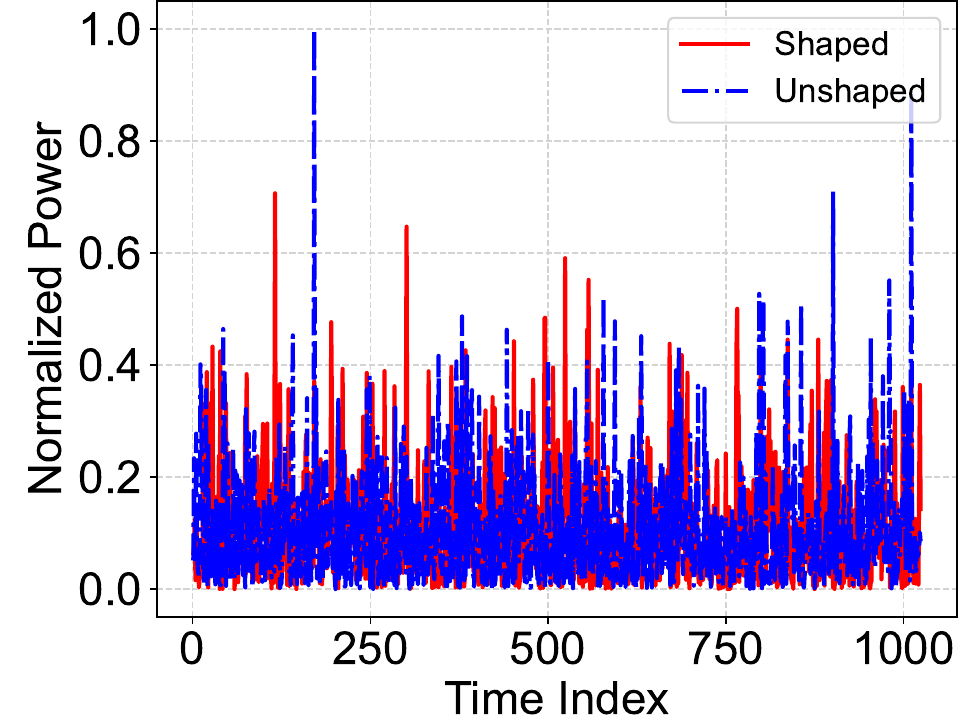}
   }
   \hspace{0.01\linewidth}
\caption{Experimental results of ISL and PAPR reduction.}
\label{fig:exp}
\end{figure}

\subsubsection{Experimental Setup}

Our testbed, as shown in Figures~\ref{fig:TX} and \ref{fig:RX}, consists of two main components. The first component is the monostatic radar system, which also functions as the communication transmitter, as depicted in Fig.~\ref{fig:TX}. The second component is the communication receiver, as illustrated in Fig.~\ref{fig:RX}. For the operation of the system, two USRP X410 devices are deployed for generating and capturing the baseband OFDM signals. Additionally, up and down converters are integrated on both the transmitter and receiver sides to enable the conversion between the baseband and the 28~GHz radio frequency band. The radar system's front-end transceiver features dedicated beamformers for transmission and reception. A horn antenna is deployed on the communication receiver side. The OFDM system operates at a full sampling rate of 200 MHz, with 1024 subcarriers allocated across the bandwidth. This configuration supports high-resolution sensing and high-data-rate communications, aligning with the requirements of advanced ISAC systems. 

\subsubsection{Experimental Results}

We first validate the trellis-shaped waveforms' ability to reduce ISL and PAPR after over-the-air transmission, accounting for real-world factors such as hardware-induced distortions, multipath effects, and environmental noise. Both shaped and unshaped OFDM signals were transmitted from the communication transmitter and captured at the communication receiver as illustrated in Fig.~\ref{fig:experiment_setting_comms}. As shown in Fig.~\ref{fig:exp1} and~\ref{fig:exp2}, the superiority of the shaped signal in terms of both ISL and PAPR is preserved even when using physical equipment on the receiver side. This result demonstrates the robustness of the proposed algorithm and its applicability to real-world implementations.

Furthermore, to directly assess the impact of ISL reduction on sensing performance, a ranging experiment was conducted using the monostatic radar setup. As depicted in Fig.~\ref{fig:experiment_setting_radar}, a metallic reflector was placed approximately 2 meters from the radar system. This distance was chosen to simulate a typical short-range sensing scenario, such as obstacle detection in indoor positioning. The range profiles were derived by computing the autocorrelation of the received signals for both shaped and unshaped waveforms.

\begin{figure}[!t]
\centering
\includegraphics[width=0.4\textwidth]{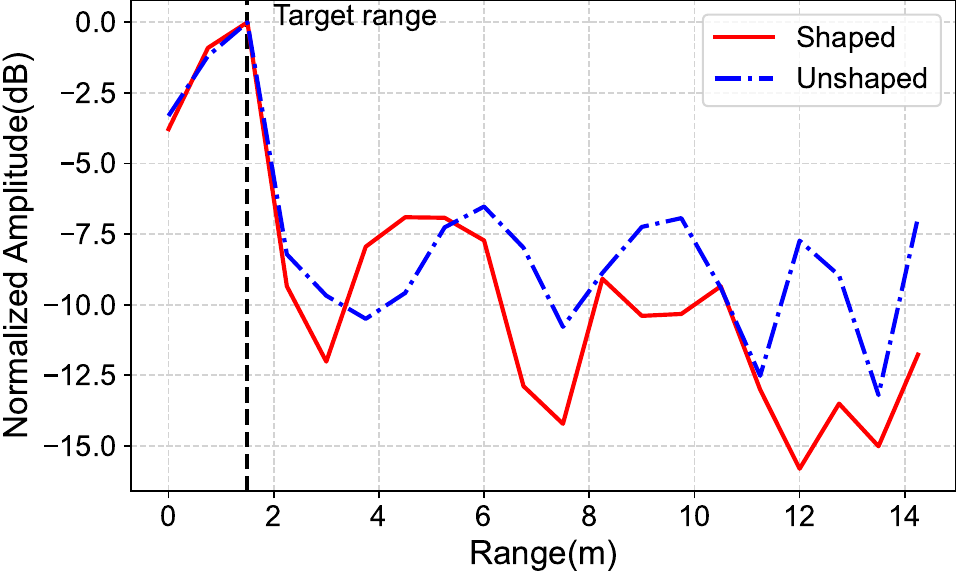}
\caption{Range profile using auto correlation}
\label{fig:Exp_range}
\end{figure}

Fig.~\ref{fig:Exp_range} presents the normalized range profiles, showcasing a clear improvement with the ISL-optimized shaped waveform. The unshaped waveform is characterized by prominent sidelobes, which pose a risk of masking weak targets and reducing detection reliability. In contrast, the shaped waveform achieves a significant reduction in sidelobe levels, producing a cleaner and more precise range profile. The mainlobe, which marks the reflector’s position, remains sharp and distinct, underscoring the success of the waveform shaping technique in enhancing sensing performance.

The hardware campaign validates the paper’s main claim: trellis shaping simultaneously cuts sidelobes and PAPR in a real mmWave ISAC link. These results, together with the extensive simulations, position the technique as a practical drop‑in upgrade for OFDM‑ISAC systems.



\section{Conclusions}\label{sec:conclusions}
In this paper, we have considered the synthesis of ISAC waveforms using trellis shaping in OFDM. Motivated by PAPR reduction in OFDM waveforms, we consider standard communication modulation and coding schemes in OFDM, and refine the waveform for better sensing performance using the trellis coding, which has been employed in PAPR reduction in OFDM communication systems. Being dual to PAPR reduction in OFDM, trellis shaping for improving the sensing performance in ISAC has been formulated using the metric of PSD variance. We have further developed the joint optimization of ISL and PAPR. The trade-off between data rate reduction and sensing performance improvement has been studied, as well as the ISL-PAPR trade-off. Numerical results show that the proposed trellis shaping approach yields a $10\sim 50\%$ reduction in ISL along with a comparable drop in PAPR. These gains have been corroborated experimentally on an mmWave SDR testbed,  confirming the practical effectiveness of the method.

\bibliographystyle{IEEEtran}
\bibliography{main}

\end{document}